\documentclass[twocolumn]{aastex7}
\usepackage{natbib}
\usepackage{needspace}
\usepackage{tabularx}
\usepackage{booktabs}
\usepackage{amsmath}
\usepackage{hyperref}
\usepackage{tablefootnote}
\usepackage{placeins}
\newcolumntype{Y}{>{\centering\arraybackslash}X}

\newcommand{\tess}{\textit{TESS}}
\newcommand{\aP}{\texttt{AltaiPony}}

\newcommand{\Gaia}{{\textit{Gaia} DR3}}

\newcommand\lowactivity{$1.99 \pm 0.07$}
\newcommand{\dif}[1]{\textmd{#1}}

\shortauthors{Capistrant et al.}

\begin{document}
\title{Stellar Flares in the TESS Light Curves of Planet-hosting M dwarfs}

\author[0000-0002-4592-8799]{Benjamin K.~Capistrant}
\email{bcapistrant@ufl.edu}
\affiliation{Department of Astronomy, University of Florida, Bryant Space Science Center, Stadium Road, Gainesville, FL 32611, USA}

\author[0000-0001-7730-2240]{Jason Dittmann}
\email{jasondittmann@ufl.edu}
\affiliation{Department of Astronomy, University of Florida, Bryant Space Science Center, Stadium Road, Gainesville, FL 32611, USA}

\begin{abstract}
    M dwarfs are magnetically active stars that frequently produce flares, which have implications for both stellar evolution and exoplanet studies. Flare occurrence rates and activity levels of M dwarfs correlate with stellar characteristics such as age, mass, and rotation period. We search TESS observations of a known active population of M dwarfs as well as a volume-limited sample of M dwarfs within 15 parsecs. We detect flares in the light curves of these stars, including 276 of 538 M dwarfs within 15 pc, and calculate cumulative flare frequency distributions (FFDs) for each star. Based on flaring behavior, we categorize stars into relatively higher and lower activity groups and fit power laws to their FFDs to compare the power law exponent ($\alpha$) across activity levels. We find $\alpha = $\lowactivity{} for the combined FFD of the lower activity M dwarfs, compared to averages of $\alpha =1.94\pm0.58$ for highly active stars with 10--100 detected flares, and $\alpha =2.03\pm0.43$ for those with $>100$ detected flares, suggesting little evolution in the power law distribution of flares as M dwarfs transition from high to low activity states. The uncertainties for the active star groups reflect the standard deviation of $\alpha$ values across individual stars within each subset. Because stellar flares and associated stellar activity complicate exoplanet observations, we also examine the subset of M dwarfs with JWST transmission spectroscopy follow-up observations in Cycles 1--3. The flares we detect for these targets are consistent with the broader 15 pc sample, providing context for interpreting planetary atmosphere retrievals from JWST spectra.
\end{abstract}

\keywords{M dwarf stars, Stellar Flares, Exoplanets, Planet Atmospheres}

\section{Introduction}
\label{sec:intro}

Stellar flares occur when a star's magnetic field reconnects, which releases a burst of radiation across the electromagnetic spectrum \citep{Kowalski_2013}. The rapid impulse of these events can enhance the optical flux by multiple orders of magnitude over timescales of minutes to seconds \citep{Haisch_1991,Schmidt_2019}, and show characteristic shapes in time-series data. \dif{Photometric surveys such as \textit{Kepler} \citep{Kepler} and \tess{}}, have therefore been useful in providing large samples of stellar flares from stars of varying properties \citep[e.g,][]{Walkowicz_2011,Shibayama2013,Davenport_2014,Hawley_2014,Aigrain_2016,Yang_2019,Gunther_2020,Feinstein_2020,Aschwanden_2021,AltaiPony,Ilin2024}. 


 The study of planetary atmospheres is at the forefront of the exoplanet field, with a particular focus on characterizing the atmospheres of Earth-like planets. These small rocky planets are more abundant around M dwarfs than around solar-type stars \citep{Dressing_2015,Mulders_2015,Hardegree-Ullman_2019,Sabotta_2021}, and are easier to detect using both the radial velocity and transit techniques, due to their relative sizes and masses with respect to their host star. The development and launch of the James Webb Space Telescope \citep[JWST;][]{jwst} has led astronomers to search for previously undetectable spectral features in small exoplanet atmospheres \citep[e.g.,][]{Lim_2023,Lustig-Yaeger_2023,May_2023,Moran_2023,Zieba_2023,Damiano_2024,Radica_2024,Alam_2024,Holmberg_2024,Wachiraphan_2024,Hu_2024,Patel_2024}. M dwarf rocky planets are reported to be the best candidates for atmospheric characterization with JWST \citep{Kempton_2018}, so investigating the physical processes of these planet-hosting stars and how they might impact their planets is a high priority. This prioritization includes the recent large-scale ($\sim 500$ hours) JWST/HST Rocky Worlds DDT Program\footnote{\url{https://outerspace.stsci.edu/display/HPR/Strategic+Exoplanet+Initiatives+with+HST+and+JWST}} proposed by \cite{Redfield_2024}, to assist in characterizing the atmospheric properties of terrestrial-mass exoplanets around M dwarfs. This program will use JWST mid-infrared secondary eclipse observations and HST observations to characterize the host star activity in the UV. As the goal of this high-priority program implies, one drawback to observing planets around these stars is how active M dwarf stars are. 

The overarching term, ``stellar activity", refers to many physical phenomena like flares, starspots, faculae, plages, and granules, all of which can have measurable effects on exoplanet transit spectroscopy observations \citep[e.g,][]{Delfosse_1998,Berdyugina_2005,Astudillo-Defru_2017, Medina_2020}. While starspots and faculae cause the largest interference in transit observations, stellar flares have been a phenomena of interest for a long time due to their explosive and therefore easily detectable signals relative to other activity processes.

Stellar activity complicates the observations of planet atmospheres \citep[e.g,][]{Sing_2011,McCullough_2014,Rackham_2017,Rackham_2018,Rackham_2023,Moran_2023,May_2023}, due to its effect on data obtained from unresolved systems where the stellar light is the primary signal. Transmission spectroscopy is one such method, which works by taking the out-of-transit, disk-integrated spectrum of the host star and subtracting it from the spectrum taken as stellar flux passes through the planet atmosphere above the limb, ideally providing an isolated spectrum of the planet's atmosphere \citep{Seager_2000,Brown_2001}. The multi-wavelength study of a planet transit can then show absorption or scattering of light in their atmospheres based on the difference in the observed exoplanet size at different wavelengths. Transmission spectroscopy has become a powerful tool for characterizing exoplanetary atmospheres, enabling detections of molecular features and atmospheric compositions. Features in the atmosphere of HD 209458 b were among the first to be identified using the \textit{Hubble Space Telescope (HST)} \citep{Charbonneau_2002,Barman_2007,Désert_2008,Deming_2013}. Since then, both HST and \textit{Spitzer Space Telescope} have been used to characterize even more exoplanets, for example see \cite{Pont_2008,Pont_2013,Tinetti_2007,Tinetti_2010,Swain_2008, Agol_2010,Desert_2011,Fraine_2014,Kreidberg_2014,Sing_2016}, among others. Ground-based facilities have also made some important contributions to atmospheric studies \citep[e.g.,][]{Snellen_2008,Snellen_2010,Bean_2010,Jordan_2013,Nikolov_2016}. 

One of the dominant uncertainties in exoplanet atmospheric measurements is that stellar heterogeneities (primarily spots or faculae) and time-variable activity from the host can affect the measured transit depth. The reference light source, typically assumed to be a perfectly homogeneous stellar disk, will not necessarily be representative of the flux that is occulted during a planet's transit. This contamination from the non-uniform stellar surface is imprinted on the planet transmission spectrum, biasing the planet atmosphere retrieval. This is known as the transit light source effect \cite{Rackham_2018}, or often referred to simply as stellar contamination \citep[e.g.,][]{Zellem_2017,Apai_2018,Garcia_2022,Barclay_2021}. The signals from stellar heterogeneities can even sometimes exceed the signal from the planetary atmosphere \citep{Rackham_2023}. This can lead to inconclusive or misinterpreted results, where a specific atmosphere scenario or spectral feature can not be definitively confirmed due to an equal probability of the feature having originated from stellar heterogeneities \citep[e.g.,][]{McCullough_2014,May_2023,Damiano_2024,Barclay_2025}.

To aid in understanding the activity of rocky planet hosts, we compute the flaring rates of planet-hosting M dwarfs from the JWST General Observer Programs in Cycles 1, 2 and 3, listed in the exoplanet category of each cycle's approved programs  \footnote{\url{https://www.stsci.edu/jwst/science-execution}}. We investigate the approved programs scheduled for transit spectroscopy measurements and exclude any coronagraphic imaging M dwarf targets. To explore the context of activity of these M dwarf targets, we run the flare detection on all of the M dwarfs in the flaring sample from \cite{Gunther_2020}, as well as a larger volume-limited sample of all M dwarfs within 15 pc determined using data from Gaia Data Release 3 \citep{GaiaDR3}. Using publicly available \tess{} light curves for all of these targets, we employ the open source python package \aP{} \citep{AltaiPony, Davenport_2016} for the detection, characterization, and analysis of the stellar flares on each of these stars. We aim to provide insight on the total activity of these stars, to connect with the results of their pending or recently published JWST transmission spectrum observations.


\section{Data}
\label{sec:data}

\subsection{TESS Photometry}
\label{subsec:TESSlcs}
\tess{} observations cover a region $24^{\circ} \times 96^{\circ}$ in size, called sectors, each continuously observed for approximately 27.4 days \citep[\tess,][]{Ricker2015}. The Science Processing Operations Center (SPOC) at NASA Ames Research Center processes these data to generate a variety of information on target stars, including the light curves investigated in this study. The reduction pipeline processes raw pixels, extracting photometry and astrometry for each target and correcting for systematic errors. Target stars are then cataloged, including their corresponding characteristics and pixel data \citep{jenkinsSPOC2016}. The SPOC pipeline uses stationary aperture photometry to generate light curves, focusing on individual target files with 2-minute or 20-second cadence data rather than full frame images (FFIs). 

To analyze the M dwarf target time series and search for flaring signals, we searched the 2-minute cadence data from all available \tess{} sectors for each star. 

\begin{table*}[hbt]

\makebox[\textwidth][c]{%
\resizebox{\dimexpr\textwidth+20pt\relax}{!}{%
\hspace*{-0.3in}
\begin{tabular}{*{15}{c}}
\hline Gaia DR3 ID& TIC ID & RA & Dec & BP-RP & Gmag & Tmag & Parallax & Distance & T$_{\mathrm{eff}}$ & Logg & RUWE & Duplicate& Flare & N$_f$ \\
(1) & (2) & (3) & (4) & (5) & (6) & (7) & (8) & (9) & (10) & (11) & (12) & (13) & (14) & (15) \\
 & & (deg) & (deg) &  &  &  & (mas) & (pc) & (K) & (dex) &  &  &  &  \\
\hline
6518076316215261568 & 144458200 & 336.2744 & -47.88254 & 2.75 & 11.21 & 1.00 & 66.71 & 14.954 & 3222.60 & 4.76 & 1.02 & - & Yes & 3 \\
1138963758544545280 & 459181676 & 123.4272 & 79.301370 & 3.19 & 13.19 & 11.80 & 66.72 & 14.957 & 3034.54 & 5.08 & 1.31 & - & Yes & 23 \\
2805045756653326080 & 611582941 & 12.6395 & 24.816575 & 2.78 & 11.24 & 10.07 & 66.73 & - & - & - & 1.54 & 13886484 & Yes & 45 \\
2232966828771055744 & 367359428 & 343.1675 & 75.071314 & 3.05 & 12.23 & 10.87 & 66.76 & - & - & - & 1.34 & - & Yes & 1 \\
2371032916186181760 & 92226327 & 11.2487 & -15.274191 & 3.03 & 12.65 & 11.30 & 66.83 & - & - & - & 1.53 & - & No & - \\
3746973763029907328 & 450341297 & 202.6271 & 19.153756 & 3.11 & 13.20 & 11.82 & 66.83 & - & - & - & 2.09 & - & No & - \\
823228721067307392 & 252470070 & 150.5864 & 48.082247 & 2.03 & 9.27 & 8.25 & 66.84 & - & - & - & 0.91 & - & No & - \\
1209281375431039104 & 157086922 & 233.8299 & 17.712292 & 2.44 & 11.35 & 10.18 & 66.88 & 14.981 & 3187.50 & 4.41 & 1.46 & - & No & - \\
4975284381907834752 & 616257329 & 11.4358 & -47.548499 & 2.83 & 12.20 & 10.80 & 66.89 & - & - & - & 2.89 & - & No & - \\
1923896725839633024 & 352503796 & 352.3502 & 41.468346 & 2.61 & 10.71 & 9.48& 66.91 & 14.888 & 3097.02 & 4.09 & 1.19 & - & Yes & 95 \\
\hline
\end{tabular}
}}

\caption{The first 10 M dwarfs in the 15 pc volume-limited sample from Gaia DR3, with identifiers and stellar parameters. The full machine-readable table of 705 stars will be made available online. Columns:
(1) Gaia DR3 source ID; 
(2) TESS Input Catalog (TIC) ID; 
(3) Right Ascension (RA) in degrees (J2000); 
(4) Declination (Dec) in degrees (J2000); 
(5) Gaia BP$-$RP color; 
(6) Gaia DR3 magnitude; 
(7) TESS magnitude; 
(8) Parallax from Gaia DR3 in milliarcseconds (mas); 
(9) Distance in parsecs (pc), inferred from Gaia DR3 parallax; 
(10) Effective temperature ($T_{\mathrm{eff}}$) in Kelvin (K); 
(11) Log of surface gravity (cm/s$^2$); 
(12) Gaia Renormalized Unit Weight Error (RUWE); 
(13) Flag for potential duplicated TIC entry (if applicable); 
(14) Flag indicating flare detection; 
(15) Number of detected flares ($N_f$).
\\\textbf{\textit{Note:}} Flare detection and statistical analysis were restricted to stars with RUWE$<1.4$ in order to reduce contamination from unresolved binaries. The version presented here, as well as the full machine-readable table, includes all stars and retains the RUWE values to enable user-defined sample selection.}
\label{tab:15pc}
\end{table*}
\subsection{Volume-limited 15 pc M dwarf Sample from Gaia DR3}
\cite{Medina_2020} performed a flare survey on the \tess{} data of a volume-complete sample of 419 main-sequence stars with masses between 0.1 and 0.3 $M_\odot$ at distances less than 15 pc \dif{determined by \cite{Winters_2021}, using data from the Gaia second data release \citep[DR2;][]{GaiaDR2, Lindegren_2018}. This upper mass limit was selected to focus only mid- to late-type M dwarfs, defining the limit for when stars become fully convective \citep{Dorman_1989,Chabrier_1997}}. \cite{Medina_2020} found that the majority of this sample did not show a single flare during the first 13 \tess{} Sectors (year 1 survey), in agreement with the findings of \cite{Feinstein_2020,Feinstein_2022}, which found fewer yet more energetic flares than other main-sequence stars. While their sample accounts for mid- to late-type M dwarfs in this time period, here we expand this sample to account for the full M dwarf mass regime ($0.08 \leq M_\odot \leq 0.57$), and more recent \tess{} Sectors (up to Sector 72).

\dif{Following the methods of \cite{Winters_2021},} we select a similar, volume-limited sample of M dwarfs within 15 pc using the Gaia third data release \citep[DR3;][]{GaiaDR3}. We search the Gaia DR3 catalog via \texttt{Astroquery}, selecting stars with parallaxes greater than 66.667 mas (distances $< 15$ pc) and limiting color to M dwarf spectral types, $1.84<(B_p - R_p)<4.86$ \citep{Pecaut_2013}\footnote{\url{https://www.pas.rochester.edu/~emamajek/EEM_dwarf_UBVIJHK_colors_Teff.txt}}. We constrain our query to search for stars with G magnitudes less than 16 mag to eliminate any false stars or artifacts \citep{Lindegren_2018,Arenou_2018,Winters_2021}. After identifying the M dwarfs within 15 pc from Gaia DR3, which we cross-match TIC IDs using the Mikulski Archive for Space Telescopes \citep[MAST;][]{MAST} catalog, giving us a total of 705 targets with available \tess{} data. \dif{Table \ref{tab:15pc}} includes the identifiers as well as some stellar parameters from the Gaia DR3 catalog.

\subsection{Günther M dwarf sample}
\label{sec:GunthSamp}
\cite{Gunther_2020} created a catalog of stellar flares, after searching all 24,809 stars with 2-minute cadence data observed in the first two sectors of \tess{} 
They found flares in the light curves of 1228 of these stars, 685 of which are M dwarfs, determined by effective temperature limits ($2285<\dif T_\mathrm{\dif{eff}}<3905$) from \cite{Pecaut_2013}. Their methodology for finding flares precedes \aP{}, though both detection methods rely on outlier heuristics. 




\subsection{Accounting for Binaries and Duplicates}

We apply additional constraints to both samples of M dwarfs to limit cases of binaries. We exclude stars with Gaia renormalized unit weight error (RUWE\footnote{\url{https://gea.esac.esa.int/archive/documentation/GDR2/Gaia_archive/chap_datamodel/sec_dm_main_tables/ssec_dm_ruwe.html}}) values greater than 1.4, as values greater than 1 can indicate the presence of an unresolved binary companion \citep[e.g.,][]{Berger_2020,Ziegler_2019}.
While this cut may not exclude most binaries \citep{Wood_2021}, RUWE is more sensitive to close binaries, which we primarily want to identify. We aim to reduce the number of cases where the origin of the flaring activity cannot be disentangled from the target star or a potential companion in the \tess{} data. We leave any M dwarfs with no reported RUWE value in the Gaia catalogs. We additionally remove 12 stars with either an eclipsing binary, or spectroscopic binary designation listed in the Simbad database \citep{simbad}, as eclipsing binaries tend to have higher flare rates than in the single star scenario \citep{Huang_2020,Yang_2023}. RUWE values for all stars in the 15 pc sample are included in \dif{Table \ref{tab:15pc}}.

We attempt to identify and remove duplicate \dif{TIC ID}s to account for cases of blended stars or stars with multiple labels whenever possible. We search for these by identifying stars with identical \tess{} observation times and then confirming their proximity, by the duplicate designation from MAST, and searching Simbad by object id and noting any with multiple TICs listed. \cite{Gunther_2020} also identified potential duplicate \dif{TIC ID}s in their catalog, so we include these along with those we determined in our final sample with the alternate \dif{TIC ID} listed in the Duplicate column (13) of \dif{Table \ref{tab:15pc}}.
 
\subsection{JWST M dwarf transmission spectroscopy targets}
\label{sec:targets}
We choose the sample of M dwarfs from the JWST GO Cycles 1, 2, and 3 exoplanet programs with publicly available \tess{} observations. Each M dwarf target in our sample has at least one sector of 2-minute cadence data reduced by the SPOC pipeline. Table \ref{tab:sectors} includes all of the targets in this sample, their \tess{} Input Catalog identifier (TIC ID), the \tess{} sectors they have been observed in, the total observation time for each target in days, the name or identifier they are referred to in their JWST program abstracts included under the ID column, as well as the JWST GO Cycle(s) they are approved to be observed in.

\begin{table*}[h!]
    \centering
    \caption{The JWST M dwarf transmission spectroscopy targets with available \tess{} SPOC light curves. The first column gives the \tess{} Input Catalog (TIC) ID of each star, and the \textbf{ID} column gives the name of the system that was provided in their JWST program abstracts. The middle column lists the \tess{} sectors the target has been observed in, followed by a column listing the total time it has been observed by \tess{} in days. 
    The \textbf{JWST Cycle} column indicates which GO cycle the target has an approved program (through Cycle 3).}
    \begin{tabular}{|*{5}{c|}}
    %
    \hline \textbf{TIC ID} & \textbf{ID} & \centering{\textbf{Available \textit{TESS} Sectors}} & \centering{\textbf{Total \textit{TESS}}} & \textbf{JWST Cycle} \\
     &  &  & \textbf{Observation Time (days)} &  \\ \hline 
        369327947 & LHS 475 & 12, 13, 27, 39, 66, 67 & 155.691 & 1 \\ \hline 
        101955023 & GJ 1132 & 9, 10, 36, 63 & 97.368 & 1 \\ \hline 
        92226327  & LHS 1140 & 3, 30 & 46.196 & 1 \\ \hline
        410153553 & LHS 3844 & 1, 27, 28, 67, 68 & 123.145 & 1 \\ \hline
        388804061 & K2-18 & 45, 46, 72 & 49.153 & 1 \\ \hline
        98796344 & LTT 1445 & 4, 31 & 51.038 & 1 \\ \hline
        34068865 & GJ 367 & 9, 35, 36, 63 & 99.772 & 1 \\ \hline
        359271092 & GJ 341 & 9, 10, 36, 37, 62, 64 & 149.337 & 1 \\ \hline
        369327947 & GJ 4102 & 12, 13, 27, 39, 66, 67 & 155.693 & 1 \\ \hline
        390651552 & GL 486/WOLF 437 & 23, 50 & 50.926 & 1,3 \\ \hline
        234994474 & L 168-9 & 1, 28, 68 & 73.961 & 1 \\ \hline
        306996324 & TOI-776 & 10, 37, 63 & 75.414 & 1 \\ \hline
        413248763 & GJ 357 & 8, 35, 62 & 72.938 & 1, 2 \\ \hline
        307210830 & L 98-59 & 2, 5, 8, 9, 10, 11,  & 528.346 & 1, 2 \\ 
        & &  12, 28, 29, 32, 35, 36, 37, 38, & & \\ 
        & & 39, 61, 62, 63, 64, 65, 69 & & \\ \hline
        278892590  & TRAPPIST-1 & 70 & 23.809 & 1, 2, 3 \\ \hline \hline
        262530407 & GJ 3090 & 2, 3, 29, 69 & 94.556 & 2 \\ \hline 
        172370679 & TOI-1899 & 14, 15, 41, 54, 55 & 129.857 & 2 \\ \hline
        28900646  & TOI-1685 & 19, 59 & 50.151 & 2 \\ \hline
        259377017 & TOI-270 & 3, 4, 5, 30, 32 & 122.919 & 2 \\ \hline
        439867639 & TOI-2445 & 4, 31 & 51.026 & 2 \\ \hline
        243641947 & TOI-3235 & 64 & 26.917 & 2 \\ \hline
        452866790 & GJ 3473 & 7, 34, 61 & 73.343 & 2 \\ \hline
        307809773 & HD 260655 & 43, 44, 45, 71, 72 & 121.194 & 2 \\ \hline
        36724087  & LTT 3780 & 9, 35, 62 & 72.740 & 2 \\ \hline
        243185500 & TOI-1468 & 17, 42, 43, 57 & 99.962 & 2 \\ \hline
        259377017 & L 231-32 & 3, 4, 5, 30, 32 & 122.919 & 2 \\ \hline
        447061717 & TOI-1231 & 9, 10, 36, 63 & 100.002 & 2 \\ \hline
        102076870 & TWA 27 & 10, 37 & 48.419 & 2 \\ \hline
        49178529  & TWA 28 & 9, 36 & 47.922 & 2 \\ \hline
        20182780  & TOI-3984 & 50, 51 & 50.817 & 2 \\ \hline
        445751830 & TOI-3757 & 59, 60, 73 & 74.708 & 2 \\ \hline
        33521996  & HATS-6 & 6, 32, 33 & 73.400 & 2 \\ \hline
        44737596  & HATS-75 & 4, 5, 31 & 76.371 & 2 \\ \hline 
        155867025 & TOI-3714 & 59 & 24.901 & 2 \\ \hline  
        \end{tabular}
        \label{tab:sectors}
\end{table*}

\begin{table*}[ht!]
\centering
        \begin{tabular}{|*{5}{c|}}
        \hline \textbf{TIC ID} & \textbf{ID} & \textbf{Available \textit{TESS} Sectors} & \textbf{Total \tess{}} & \textbf{JWST Cycle} \\
     &  &  & \textbf{Observation Time (days)} &  \\ \hline 
        150428135  & TOI-700 & 1, 3, 4, 5, 6, 7, 8, 9, & 710.161 & 3 \\
        & &  10, 11, 13, 27, 28, 30, 31, 33,& & \\ 
        & &  34, 35, 36, 37, 38, 61, 62, & & \\
        & &  63, 64, 65, 66, 68, 69  & & \\ \hline
        181804752 & LP 791-18 & 9, 36, 63 & 72.063 & 3 \\ \hline
        86263325 & TOI-3884 & 46,49 & 49.912 & 3 \\ \hline
        11893637 & GJ 1151 & 22, 48 & 49.316 & 3 \\ \hline
        441420236 & AU MIC & 1, 27 & 51.465 & 3 \\ \hline
        239332587 & TOI-4481 & 41, 55, 56, 75, 76 & 136.407 & 3 \\ \hline
        381979901 & TOI-1085 & 12, 13, 27, 28, 29, 30, & 587.679 & 3 \\
        & & 31, 32, 33, 39, 38, 37, 36, & & \\
        & & 35, 34, 61, 62, 63, 64, 65, & & \\
        & & 66, 67, 68, 69 & & \\ \hline
        166184428 & TOI-4336 & 38, 64 & 50.982 & 3 \\ \hline
    \end{tabular}
    
    \label{tab1continued}
\end{table*}

\section{Methods}
\label{sec:analysis}
We use the open source python package \aP{} \citep{AltaiPony,Davenport_2016} for our flare analysis of each M dwarf sample. \aP{} provides a series of tools for a statistical flare analysis such as light curve de-trending, flare detection and characterization, injection-recovery of synthetic flares, and computation of flare frequency distributions (FFDs) for individual stars as well as larger collections of stars. In this section, we describe our methodology, which largely follows the process performed by \cite{AltaiPony}. 

In the following subsections, we describe how we identify flare candidates in the 2-minute cadence SPOC reduced \tess{} light curves, and compute their equivalent durations (EDs) ($\S$ \ref{sec:detect}). Because EDs represent the relative flare energy with respect to each stars' quiescent flux \citep{Gershberg1972}, we perform injection-recovery tests of synthetic flares to show the detection sensitivity of \aP{} on \tess{} light curves of non-uniform noise levels ($\S$ \ref{sec:I-R}). The FFD module (\texttt{ffd.py}) of \aP{} allows us to then find and plot the flare frequency distribution for each JWST target, which we compare to the FFDs of the larger sample of M dwarfs from the \cite{Gunther_2020} and volume-limited 15 pc sample ($\S$ \ref{sec:FFD}). 
In this subsection, we also describe the methods we use for fitting power laws to these distributions, and the analysis of the power law behavior for stars with lower flaring compared to those that may have higher levels of activity.


\subsection{Light curve de-trending \& Flare detection}
\label{sec:detect}
Flare durations occur on time scales of minutes to hours, and rarely longer than days, whereas stellar variability typically occurs on longer time scales. Rather than manually de-trending the \tess{} light curves as \cite{AltaiPony} performed with their sample of K2 light curves \citep{K2}, we make use of the available SPOC reduced 
2 minute data for each star. \aP{} also applies some simple additional de-trending methods, which are explained in detail in \cite{AltaiPony} and \cite{IlinTESS2022}, but we briefly describe the process here. 

We first read in the SPOC reduced \tess{} data from MAST \citep{MAST} for each given star and a Savitky-Golay filter is applied to smooth each light curve \citep{Savgol1964}. This is a simplified least squares procedure, meant to flatten the light curves with a moving average. We adopt a window length of 3 hours for each continuous section within the light curve, defined as time series with no gaps longer than 2 hours. In the smoothing process, single data point outliers are iteratively clipped above 3$\sigma$, and replaced with the previous median value, until convergence to determine the quiescent flux in the light curve. The flux error from SPOC is used to estimate the noise of the light curve as well as the quiescent flux.

The criteria for searching for the residual light curves for flares is based on equations from \cite{Chang2015}. Three consecutive data points 3$\sigma$ above the quiescent flux threshold are flagged as flare candidates, and subsequent ($\sqrt{n}$) points before and after are masked to capture the slow rise and extended decay shape of each flare.


The equivalent duration (ED) of a flare is defined as the flare flux divided by the median quiescent flux, $F_0$, of the star, integrated over the flare duration \citep{Gershberg1972}
\begin{equation}
    \mathrm{ED} = \int \frac{F_{\mathrm{flare}}(t)}{F_0}dt
    \label{eq:ed}
\end{equation} 
This quantity acts as a proxy for the flare energy and is suited for comparing flaring activity on stars as it is independent of stellar calibration and distance, though it is dependent on properties like the light curve noise, sampling, space-craft systematics, and de-trending effects. It can be longer or shorter than the actual flare duration as it describes the time during which the non-flaring star releases as much energy as the flare itself.

\subsection{Synthetic Flare Injection-Recovery}
\begin{figure*}[h!]
    \centering
    \includegraphics[width=0.99\textwidth]{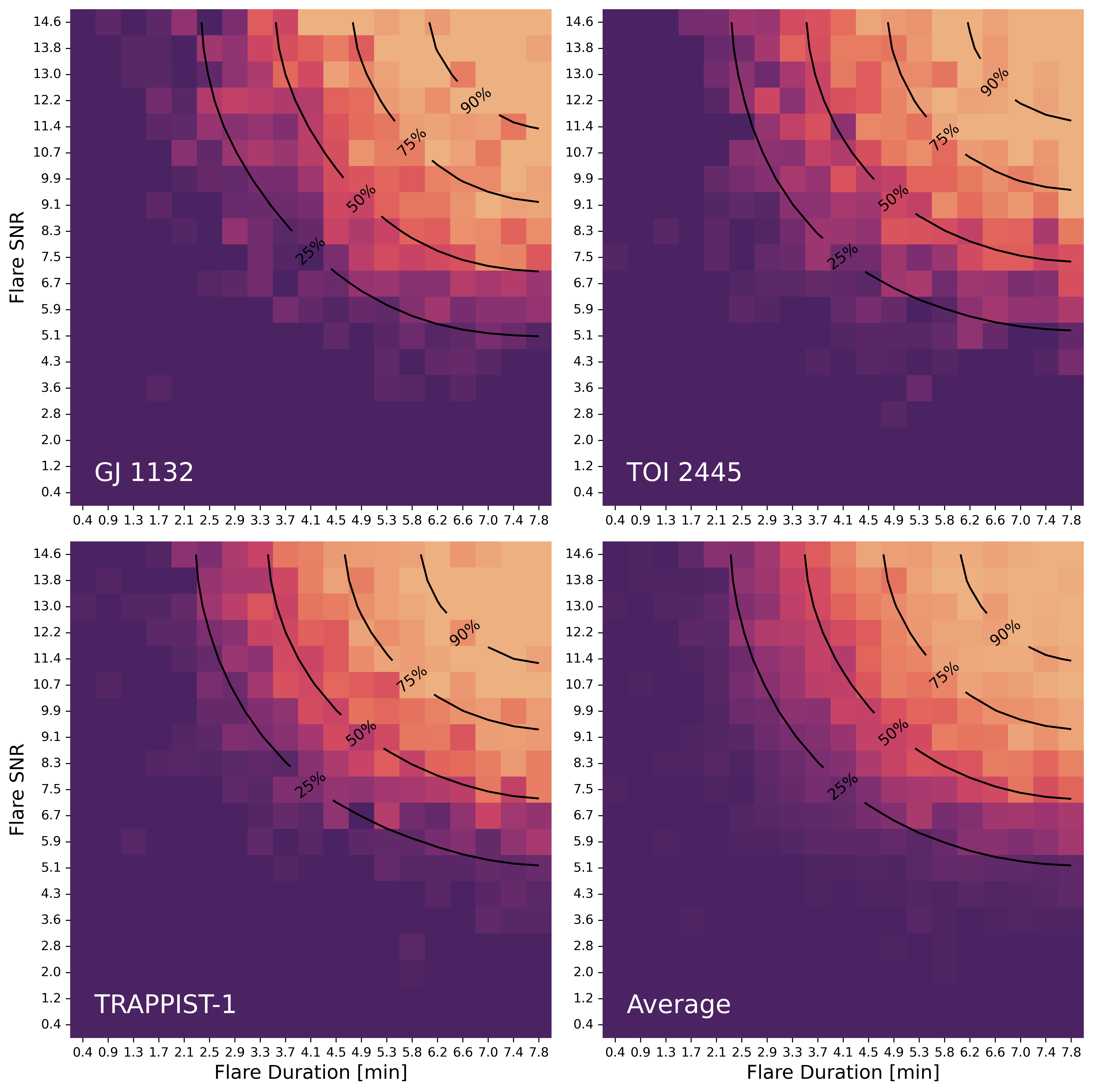}
    \caption{The recovery probability (successful \aP{} detection) of injected flares in the \tess{} light curves of GJ 1132 (top left), TOI-2445 (top right), TRAPPIST-1 (bottom left) and the average of recovered flare durations (in minutes) and flare SNR for all three stars (bottom right). From tests on approximately 6500 injected flares per star, these heatmaps show the probability that \aP{} detects a flare of a given duration and amplitude relative to the light curve noise. The color of each bin represents the recovery probability from 0 to 1, with black contours indicating the approximate curves of of 25\%, 50\%, 75\% and 90\% recovery probability. Although the noise levels within each stars' light curve varies greatly, we see consistent recovery probability of flares. }
    \label{fig:FLIR}
\end{figure*}
\label{sec:I-R}
To investigate the detection limits of pipeline for our light curves, we implement synthetic flare injection-recovery tests using the \aP{} method \texttt{sample\_flare\_recovery}. Because the flare detection relies on outliers above the inherent scatter of each light curve, injection-recovery allows us to observe the detection sensitivity of \aP{} in this sample of \tess{} light curves. This method involves iterative injection of flares into the SPOC light curves of a given star, at least one and no more than five, per continuous observing period. We implement a proxy for the signal-to-noise ratio (SNR) for a given real or synthetic flare detection, using the flare amplitude in units of relative flux, divided by the median absolute deviation (MAD) of the detrended light curve flux that is being searched. This allows us to check for a consistent detection threshold across stars who's \tess{} light curves often show varying levels of noise or scatter that may impact a direct measurement of a flare amplitude. 

We have \aP{} inject synthetic flare signals with flare durations between 12 seconds and 8 minutes, and flare amplitudes up to a SNR calculation of 15 for a given light curve. 
\aP{} then de-trends the light curve and runs the detection as described in Section \ref{sec:detect}, and repeats the process for approximately 6500 injections per light curve. We bin the injected flare SNR with the injected flare durations and calculate the probability of recovery (\aP{} detection) within each bin. 

Figure \ref{fig:FLIR} shows heat maps of the recovery probabilities for the flare injection-recovery tests we performed on the light curves of targets GJ 1132, TOI-2445, and TRAPPIST-1, which have different noise levels in their respective TESS observations. For one sector of \tess{} data for each of these stars, we find the noise in units of relative flux from MAD calculations of 0.0012, 0.0045, and 0.0093 respectively. At an SNR of 15, representing the maximum value we inject, this corresponds to a flare amplitude of $\sim 1.8 \%$, $\sim 6.7 \%$ and $\sim 14 \%$ in relative flux units, confirming that we are sensitive to detecting smaller flares in light curves with less scatter. The bottom right panel shows the averaged recovery probabilities across all three stars for flares of these sizes, also indicating the consistency of \aP{} detection and a consistent dependence on the light curve scatter. The black contours are overlaid on each heatmap to show the values of flare SNR and flare duration we are sensitive to with 90\% recovery probability, 75\%, 50\% and 25\%, as labeled.

We find the lower limit of the \aP{} flare detection is approximately 2 minutes in flare duration, and a flare SNR of 4-5 based on its amplitude over the light curve MAD. These recoveries show that our detection sensitivity finds lower amplitude flares in quieter light curves, as is a well known drawback of outlier heuristic detection methods \citep{Feinstein2020a,AltaiPony,Vida_2021}, but this limitation remains consistent across our large sample of light curves.

\subsection{Flare frequency distributions \& Power laws}
\label{sec:FFD}
The flare frequency distribution is derived from the frequency of detected flaring events over the total observation time for a given target which we plot against the equivalent duration, or flare energy. The cumulative flare frequency is calculated based on the ordered array of flare equivalent durations, or ED$_n$ for the $n$th flare. The frequency of a flare with ED$>$ED$_n$ is $n$ (the number of flares larger than ED$_n$) divided by the total observing time \citep{Hunt-Walker_2012}. \aP{} then requires the total observing time of each target to output the value for cumulative flare frequency in $\nu$/days, thus allowing for comparison between stars observed over different time intervals. We calculate the total observing time for each target as each light curve is searched for flares, by summing over the time stamps excluding those falling in data gaps, since no flares can be detected there. Data gaps can occur in \tess{} light curves due to data downlink periods, momentum dumps, observing sector boundaries, pointing adjustments and calibrations, as well as other sources of interference during its continuous observing. 

\cite{Gershberg1972} and \cite{Lacy_1976} were among the first studies to measure the power law relation between flare frequency and flare energy in their statistical analyses of UV Cet, YZ CMi, EV Lac, and AD Leo flares. This has since been observed for stars showing flares of varying sizes and energies such as superflares, solar flares, microflares, and nanoflares \citep{Shibayama2013}. The power law takes the form
\begin{equation}
    f(>E) = \frac{\beta}{\alpha-1} E^{-\alpha+1}
    \label{eqn:pl}
\end{equation} 
for flares above a given energy E or equivalently ED. The observed cumulative flaring frequency can vary for each star, and studies have investigated correlations with stellar parameters such as mass, effective temperature, spectral type and stellar rotation \citep[e.g,][]{Lin_2019,AltaiPony,IlinTESS2022,Feinstein_2022,Ilin2024}. These studies generally find more frequent flaring for younger, faster rotating stars, though correlation with spectral type is still debated and seemingly limited by flare detection completeness. Studies of optical flare frequency distributions across stars of all spectral types have found some variation in calculated power law exponents, $\alpha$ (also commonly referred to as the power law index or slope), but it is typically around $\alpha \approx 2$ \citep{Lacy_1976,Shakhovskaia_1989,Hilton_2011,Shibayama2013,Hawley_2014,Lurie_2015,Chang2015,Lin_2019,Yang_2019,Aschwanden_2021,Feinstein_2022,Yang_2023,Lin_2024}. See Figure 13 of \cite{AltaiPony} for a comparison of derived values of $\alpha$ from the studies predating it in which power laws were fit to FFDs of flares in the optical regime. The authors show the distribution of calculated $\alpha$ values from those studies, which range from $\sim 1.4-2.6$, but with a peak at $\alpha = 2$. They conclude that it's too difficult to determine if this spread in values is due to a physical effect or skewed by the variation in power law fitting methods, or a lack of reliable uncertainty estimates between studies. \cite{Aschwanden_2021} performed a study of FFDs of all Kepler stars, and compared $\alpha$ values from a variety of sources pre-dating it, reporting a similar range of values (see their Tables 1 and 2) and an average of $\alpha = 1.93\pm0.4$. For only M-type stars, they report a value of $\alpha =1.99\pm0.35$ from their power law fits. We discuss and compare our findings with these, as well as more recent works, in our discussion section ($\S$ \ref{sec:ffdPLs}).

\begin{figure*}[h!]
    \centering
    \includegraphics[width=0.48\textwidth]{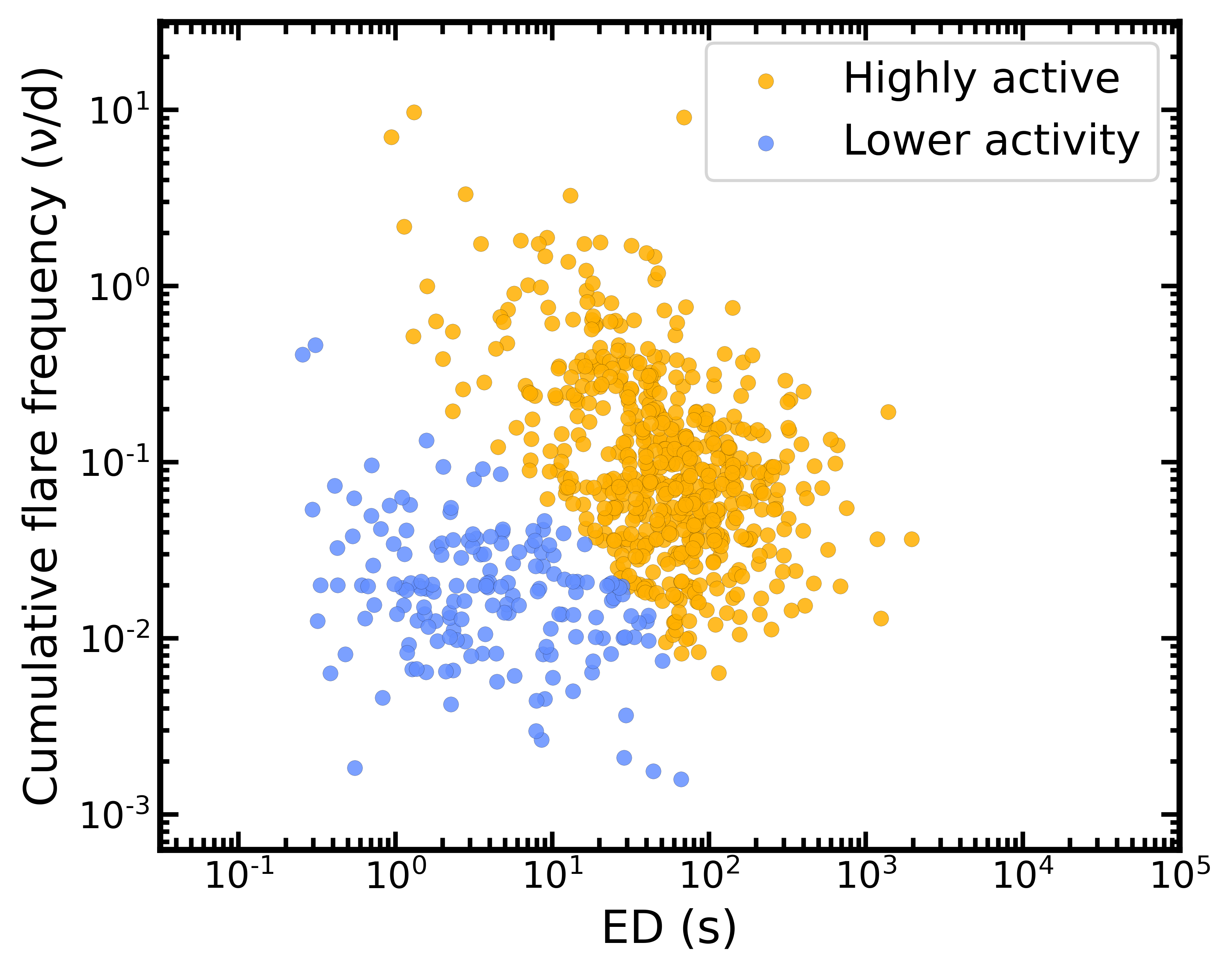}
    \includegraphics[width=0.48\textwidth]{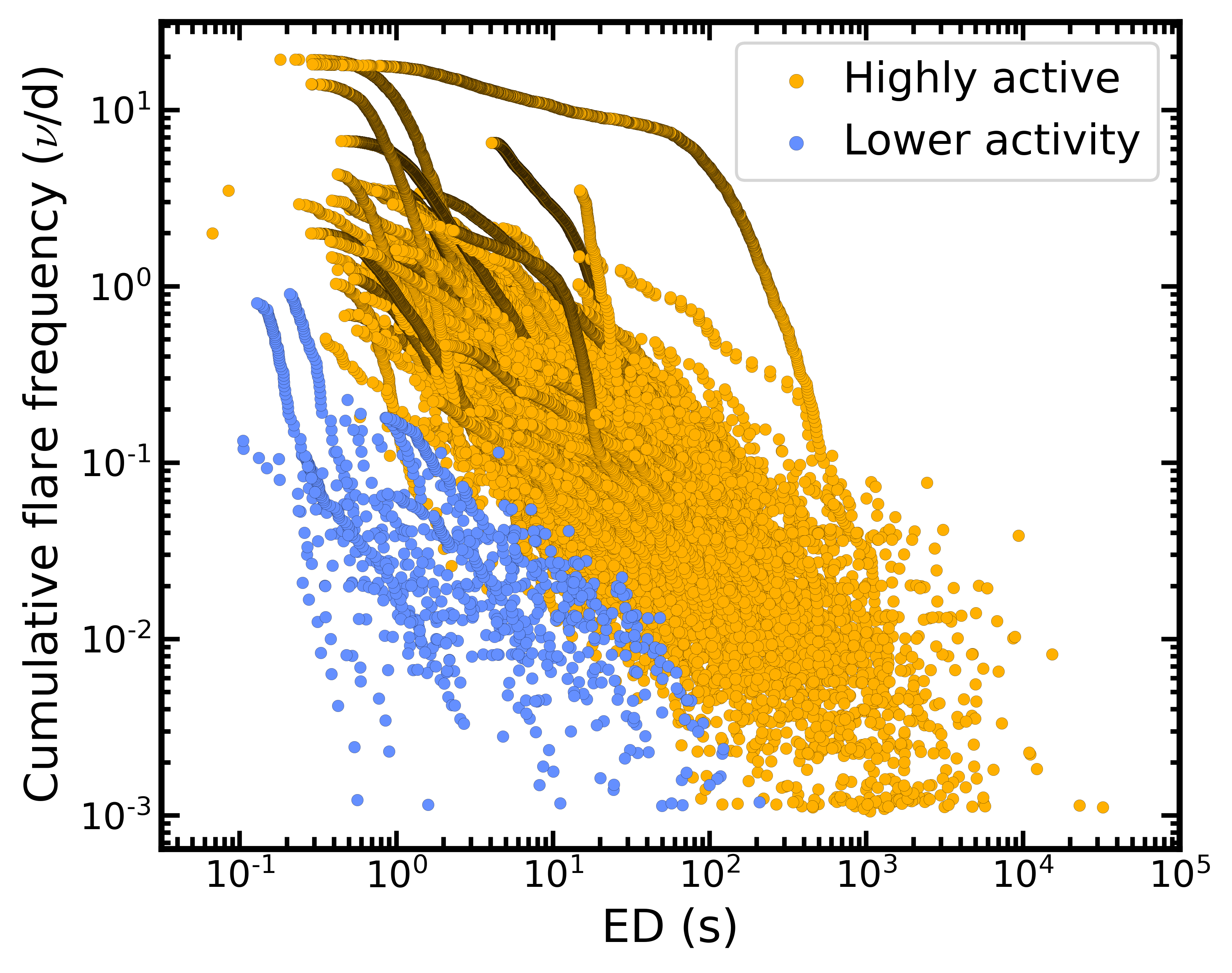}
    \caption{Clustering of stellar flare activity. \emph{Left panel:} Each point represents one star, plotted in log‑log space of mean equivalent duration (ED) in seconds, versus cumulative flare frequency ($\nu$/d). Means are computed after excluding the lowest 20\% of flares by ED to reduce the effects of low-energy flare detection incompleteness. A two‑component Gaussian Mixture Model (GMM) separates the stars into two clusters: lower activity (blue) and highly active (gold). \emph{Right panel:} Full flare frequency distributions (FFDs) for each star, colored by their GMM-assigned label.}
    \label{fig:PLsplit}
\end{figure*}

\begin{figure*}[!hbp]
    \centering
    \includegraphics[width=0.495\textwidth]{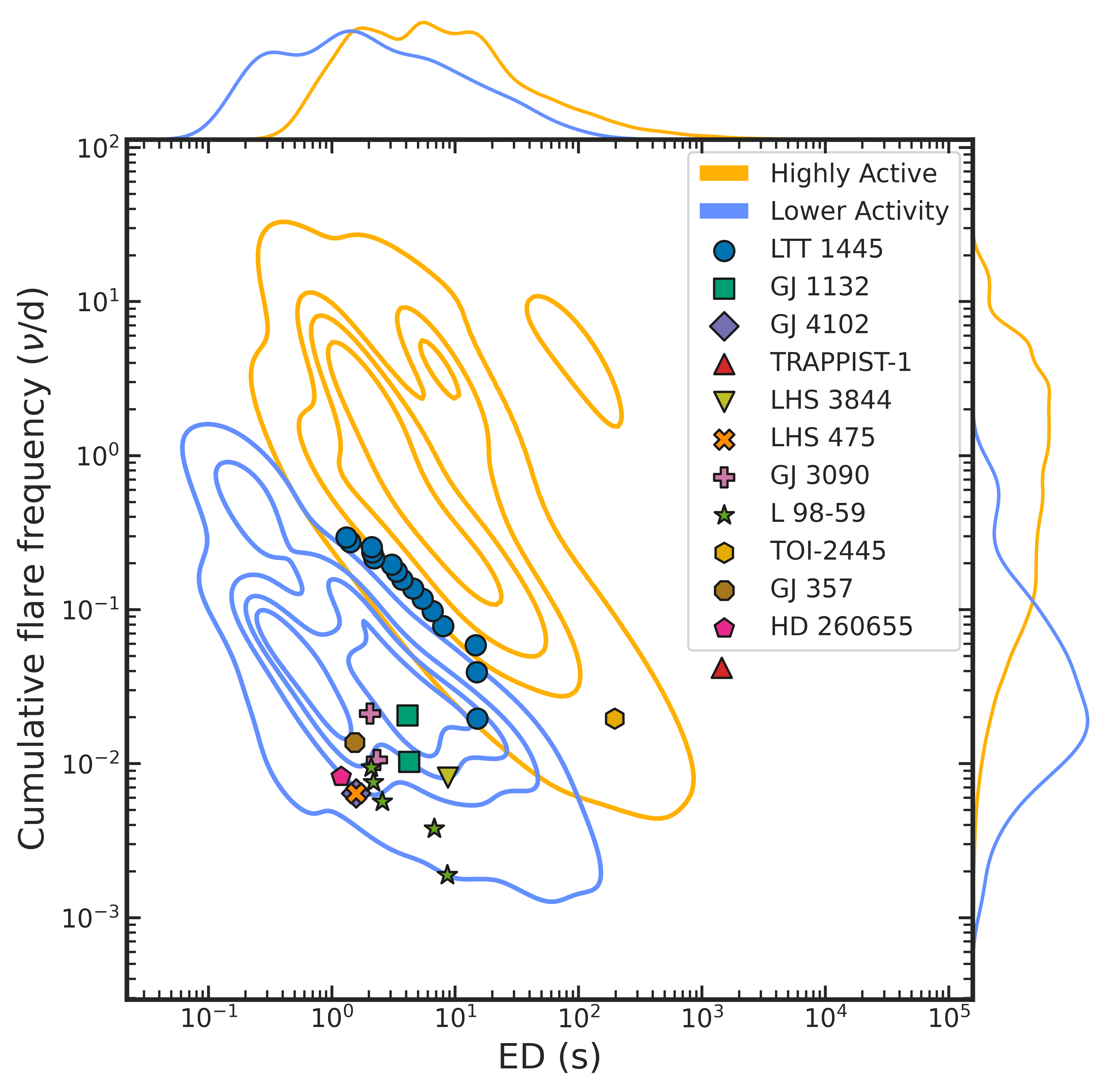}
    \includegraphics[width=0.495\textwidth]{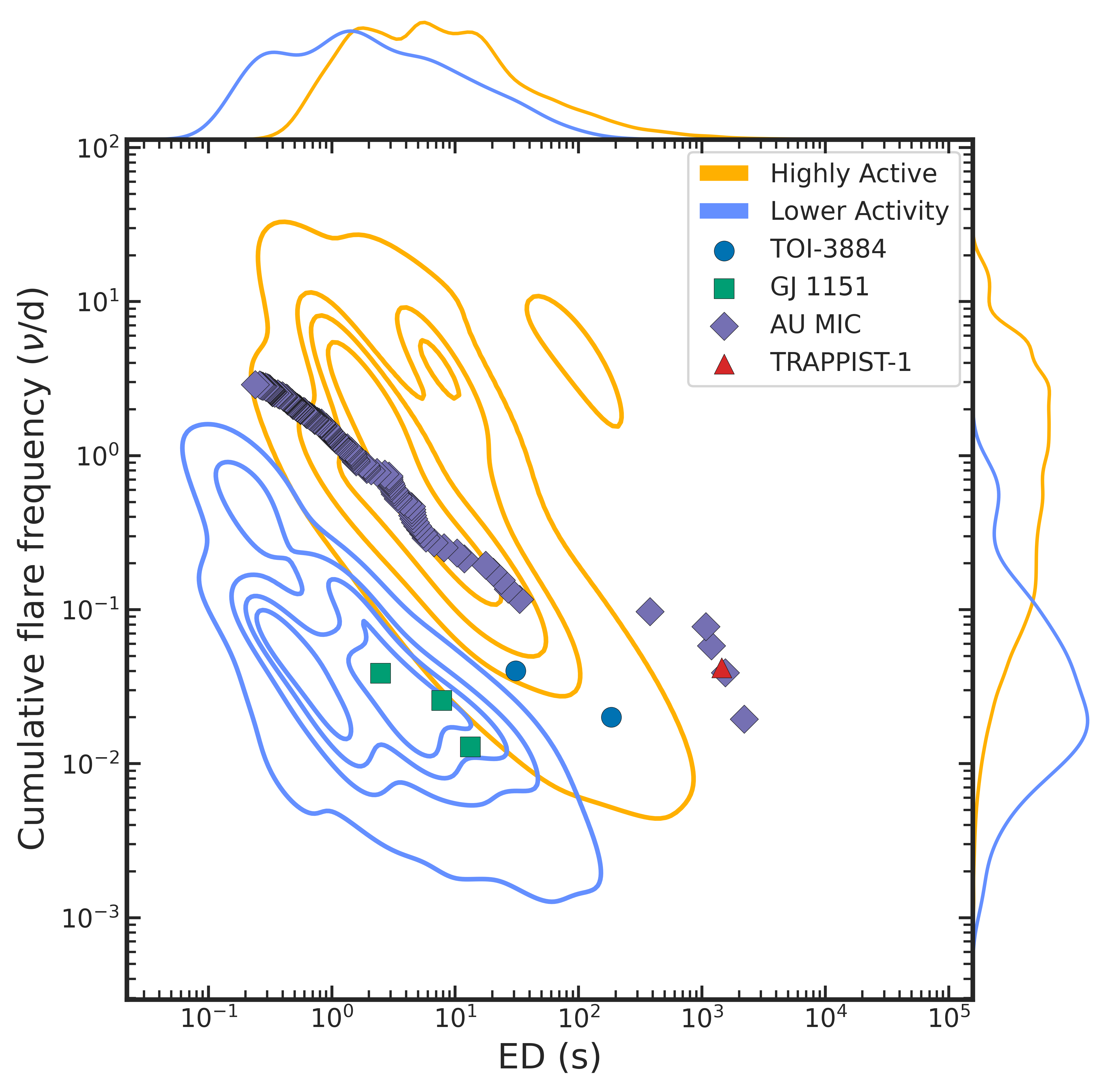}
    \caption{Flare frequency distributions for all M dwarf targets from detected flares in their SPOC light curves. In both panels, the blue and gold contours represent the flares from the combined \cite{Gunther_2020} and 15 pc samples, split by \dif{relative activity, based on the GMM clustering} shown in Figure \ref{fig:PLsplit}. The density plots on the outer edges represent the normalized distributions for flare equivalent duration in seconds (ED) on the x-axes, and cumulative flare frequency on the y-axes. Each scatter point represents a flare we detect for the different JWST targets which are differentiated shape and color. The left panel contains our computed FFDs of 11 M dwarf transmission spectroscopy targets from JWST GO cycles 1 and 2 and the right panel shows FFDs we computed for 4 M dwarf targets in Cycle 3. The Cycle 1 and 2 targets largely fall in the lower activity regions aside from TOI-2445 and TRAPPIST-1, whereas the Cycle 3 targets largely fall in the highly active region with the exception of GJ 1151. Note that TRAPPIST-1 has been observed in each JWST Cycle, which is why it is included in both panels. We expect increased stellar contamination in planet atmosphere transmission spectra obtained from more active stars.  
    }
    \label{fig:FFD}
\end{figure*}

\dif{We compute cumulative FFDs for each star in both the \cite{Gunther_2020} and the Gaia DR3 15 pc samples. To investigate the power law behavior of stars of varying activity, we categorize each star by either relative higher or lower activity using Gaussian Mixture Models (GMM) clustering \citep{Reynolds_2009}. Each star was classified based on its mean equivalent duration and frequency values. To mitigate some bias from detection incompleteness at low energies, we excluded the bottom 20\% of flares by ED before computing these, to reduce the influence of the lowest-energy flares on the mean values. We applied the two-component GMM clustering on these values in log-space, allowing for a full covariance, which provides probabilistic cluster assignments for each star. In the left panel of Figure \ref{fig:PLsplit}, the mean values for each star are plotted and colored by cluster membership: blue for the ``lower activity" and gold for what we label ``highly active" stars. In the right panel we show the full cumulative FFDs for each star in these samples, still colored based on the cluster membership assigned by the GMM. \cite{AltaiPony} suggests that relative flaring activity is best described by the $\beta$ parameter, or power law intercept, in Eq. \ref{eqn:pl} due to the definition of ED being relative to the quiescent stellar flux. We select GMM clustering due to its probabilistic handling of overlapping data, and to select oblong clusters that visually appear to separate by $\beta$ parameter rather than other clustering algorithms that would naturally separate by average ED. We recognize the split may be imperfect, but the clustering encompasses relatively inactive stars in this sample, where the greatest ED flare will also be $\lesssim 100$s.}

Figure \ref{fig:FFD} displays the flares from these stars as contours and normalized histograms on the figure edges, with their color indicating the lower activity or highly active categories (colored blue and gold respectively). The FFDs of the JWST transmission spectroscopy M dwarf targets are plotted over these contours as scatter points of different colors and shapes, corresponding to each target star we detected flares. The left panel includes flaring M dwarf targets from JWST GO cycles 1 and 2, and the left panel includes the targets from the list of approved programs for GO cycle 3. The flare detected in the single \tess{} observation for TRAPPIST-1 is included in both panels, as it has been observed in each JWST Cycle to date, and is known to be a highly active M dwarf \dif{\citep{Vida_2017,Howard_2023}}, in agreement with our detection and classification. 

Highly flaring M dwarf AD Leo shows up in the volume-limited 15 pc sample, and its FFD can be seen in Figure \ref{fig:PLsplit}, falling above the most highly active stars with many low-energy flares occurring at a high frequency before a turning point around ED $\sim10^2$ seconds, followed by a steeper power law. Its high level of flaring has previously been observed and presented in literature \citep{Hawley_2003,Hunt-Walker_2012,Bai_2023}. Due to the broken power law shape of its FFD and the atypical behavior of its flaring, we exclude these flares from the highly active (gold) contours of Figure \ref{fig:FFD} to better compare with the JWST M dwarfs, which show more typical flaring behavior.

We implement a fitting routine to investigate the power law behavior of the M dwarf FFDs \dif{for the relative levels of activity that we determined}. In the FFD module of \aP{}, the \texttt{multiple\_star} argument is recommended by the authors when analyzing samples of less than 100-200 flares. Because many of the FFDs for stars in the lower activity category consist of fewer flares, we create a multiple-star FFD to compare this sample with the highly active FFDs. We use the \textit{Easy Differential-Evolution Markov Chain Monte Carlo algorithm}, \texttt{edmcmc} \citep{vanderburgedmcmc}\footnote{\url{https://github.com/avanderburg/edmcmc}}, an open source Python MCMC routine. We employ the algorithm with Equation \ref{eqn:pl} to fit the power law parameters ($\alpha$ and $\beta$) to the combined lower activity FFD. 
We use the literature value of 2 for the initial starting position for the $\alpha$ parameter, and a starting position for $\beta = 0.015$ based on a by-eye inspection of the distribution. We include only one uniform prior that $\beta$ must be non-negative. We run the MCMC for $10^4$ steps, with 200 walkers (number of individual chains), and 2000 burn-in steps (initial steps of the chain to discard before the algorithm approaches equilibrium). We employ the \cite{GelmanRubin} statistic to ensure the algorithm has converged to the true posterior distribution.

\begin{figure*}[!ht]
    \centering
    \includegraphics[width=0.45\textwidth]{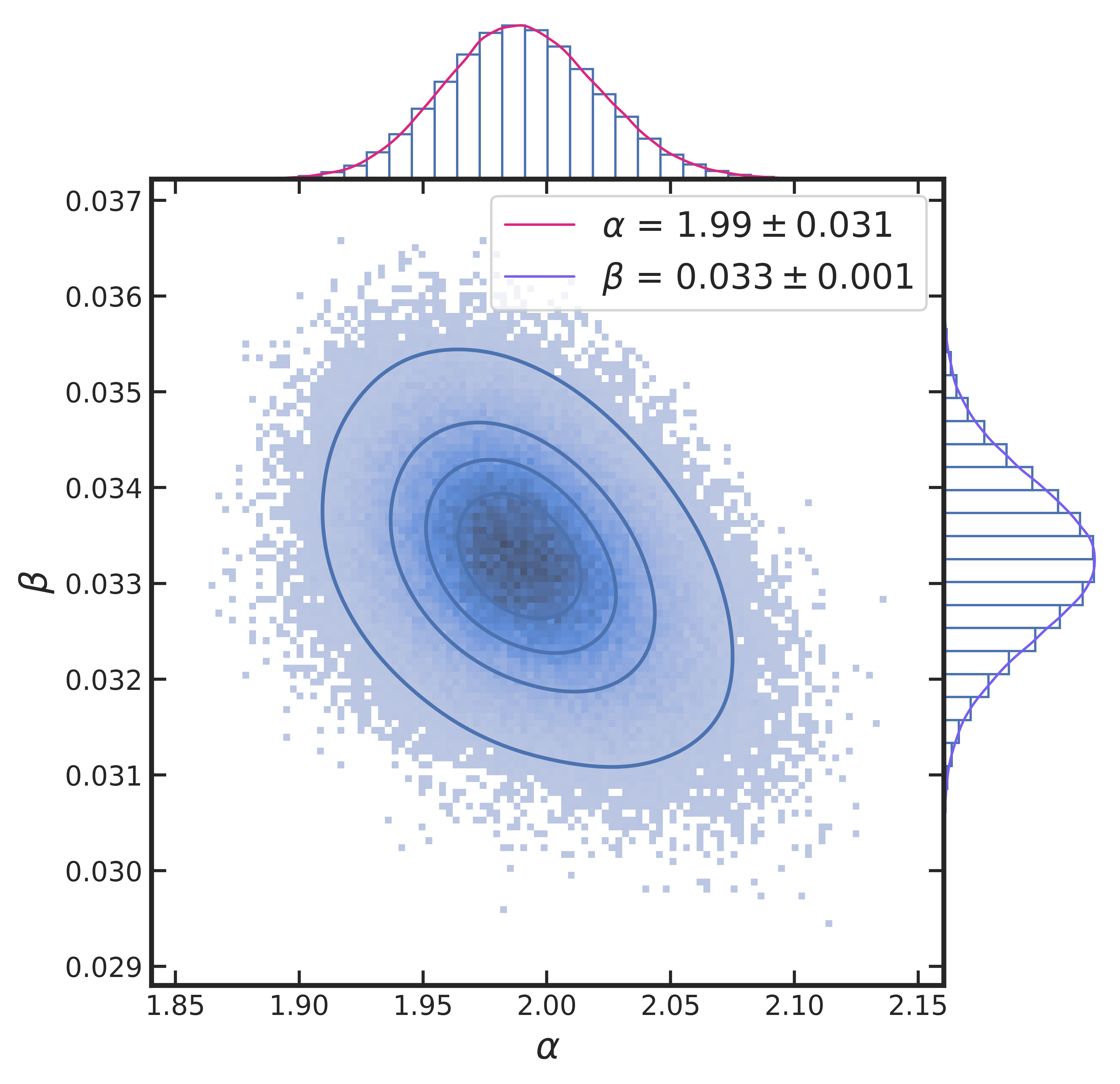}
    \includegraphics[width=0.45\textwidth]{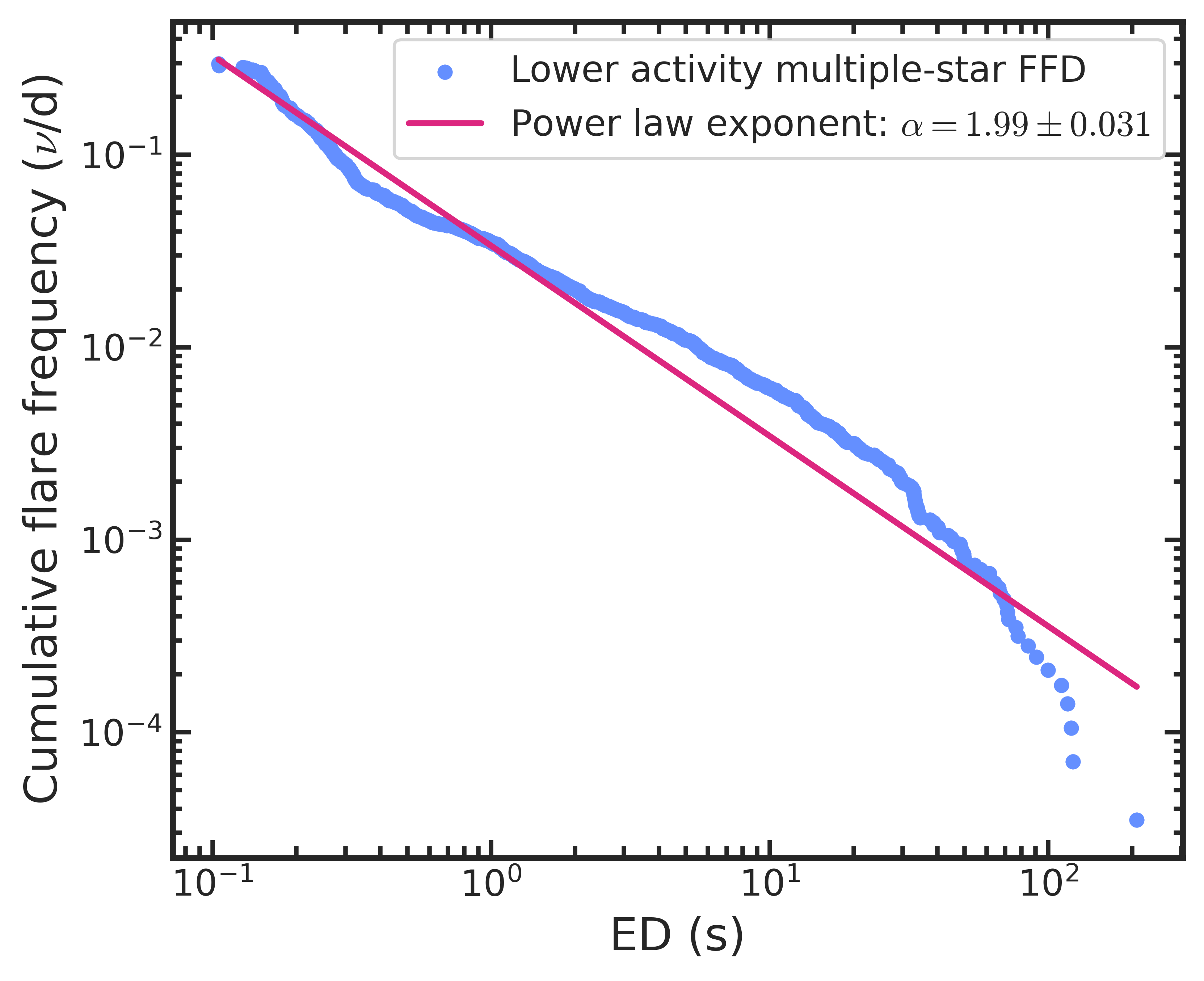} 
    \caption{The power law fit to the flares of the lower activity stars in our M dwarf samples (\dif{170} stars), indicated by the blue points in Figure \ref{fig:PLsplit}. We treat these stars as one distribution (containing \dif{858} flares) and use the \texttt{multiple\_star} argument of the \aP{} FFD module to average each portion by the number of stars that contribute to it. We employ \texttt{edmcmc} to find optimal power law parameters $\alpha$ and $\beta$, with the posteriors of the algorithm's search shown in the left panel. The right panel shows the flare frequency distribution in blue with the power law (Equation \ref{eqn:pl}) fit to it using the optimal parameters from the MCMC run in pink. Based on the posteriors, this distribution is best fit with a power law exponent, $\alpha$, of $1.99 \pm 0.031$, and a $\beta$ parameter of $0.033 \pm 0.001$. This value for $\alpha$ is in agreement with literature findings for power laws fit to flare frequency distributions in the optical regime. See our discussion in Section \ref{sec:ffdPLs}, or Fig. 13 of \cite{AltaiPony}, Tables 1 \& 2 of \cite{Aschwanden_2021}, and more recent findings for power law exponents fit to M dwarf FFDs \citep{Yang_2023,Lin_2024} \dif{for example}. }
    \label{fig:mcmc}
\end{figure*}

Figure \ref{fig:mcmc} shows the posteriors from the MCMC run for the two power law parameters in the left panel with the resulting power law fit of the combined lower activity flare sample on the right. This fit is indicated by the pink line with well converged solutions of $1.99 \pm 0.03$ for $\alpha$, and $0.033\pm0.001$ for $\beta$.
With this power law fitting routine, we additionally compute new power law fits for each of the FFDs for the highly active stars for those with more than 10 flares identified in their light curves. We found the Gelman-Rubin statistic was often indicating convergence (values near unity) with fewer steps than our previous fit, so we decrease this to $10^3$ steps and 200 burn-in steps to fit the 445 highly active FFDs to speed up our computation time. For any MCMC runs on distributions that did not reach convergence, meaning Gelman-Rubin values greater than 1.1, we re-run the algorithm with the larger $10^4$ step size and 2000 burn-in steps. These larger runs provide well converged solutions for all of distributions that failed to converge with the shorter runs. The power law relationship is often best fit to the middle of flare distributions, but suffers from low detection efficiency at the low energy end of the distribution and becomes truncated at the higher energy end \citep{AltaiPony}. Because we see similar behavior in many of the highly active FFDs, we exclude 20\% of each stars lower energy flares to provide better starting positions for the MCMC to initialize its $\alpha$ and $\beta$ fits.

\begin{figure}
    \centering
    \includegraphics[width=0.99\columnwidth]{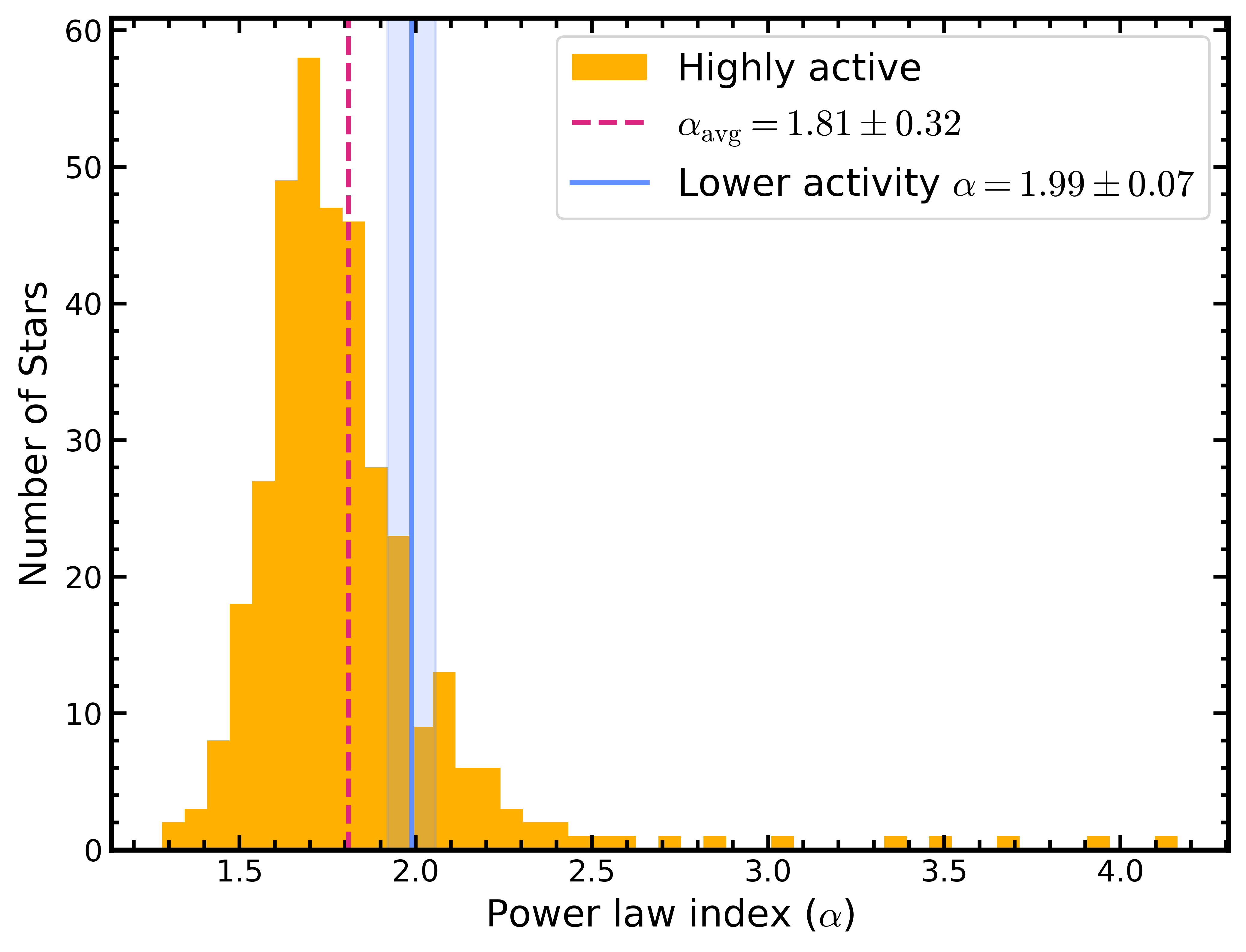}
    \caption{Histogram of the power law exponents ($\alpha$ values) fit to \dif{362} M dwarfs from both the \cite{Gunther_2020} and Gaia DR3 volume-limited 15 pc samples in the highly active category defined by Figure \ref{fig:PLsplit}. The gold distribution represents exponents fit to highly active stars with more than 10 detected flares (362 stars), which has an average value $\alpha _{\mathrm{avg}}$ of \dif{$1.81$ and a standard deviation $\pm 0.32$}, indicated by the pink vertical line. The vertical blue line represents the $\alpha$ value best fit to the combined lower activity star FFD shown in Figure \ref{fig:mcmc}) with the shaded region representing the uncertainty estimated with Equation \ref{eqn:uncert}. The averages are consistent with each other and with literature findings for $\alpha$ values of optical flares for M dwarfs. }
    \label{fig:slopes}
\end{figure}

\begin{figure}
    \centering
    \includegraphics[width=0.99\columnwidth]{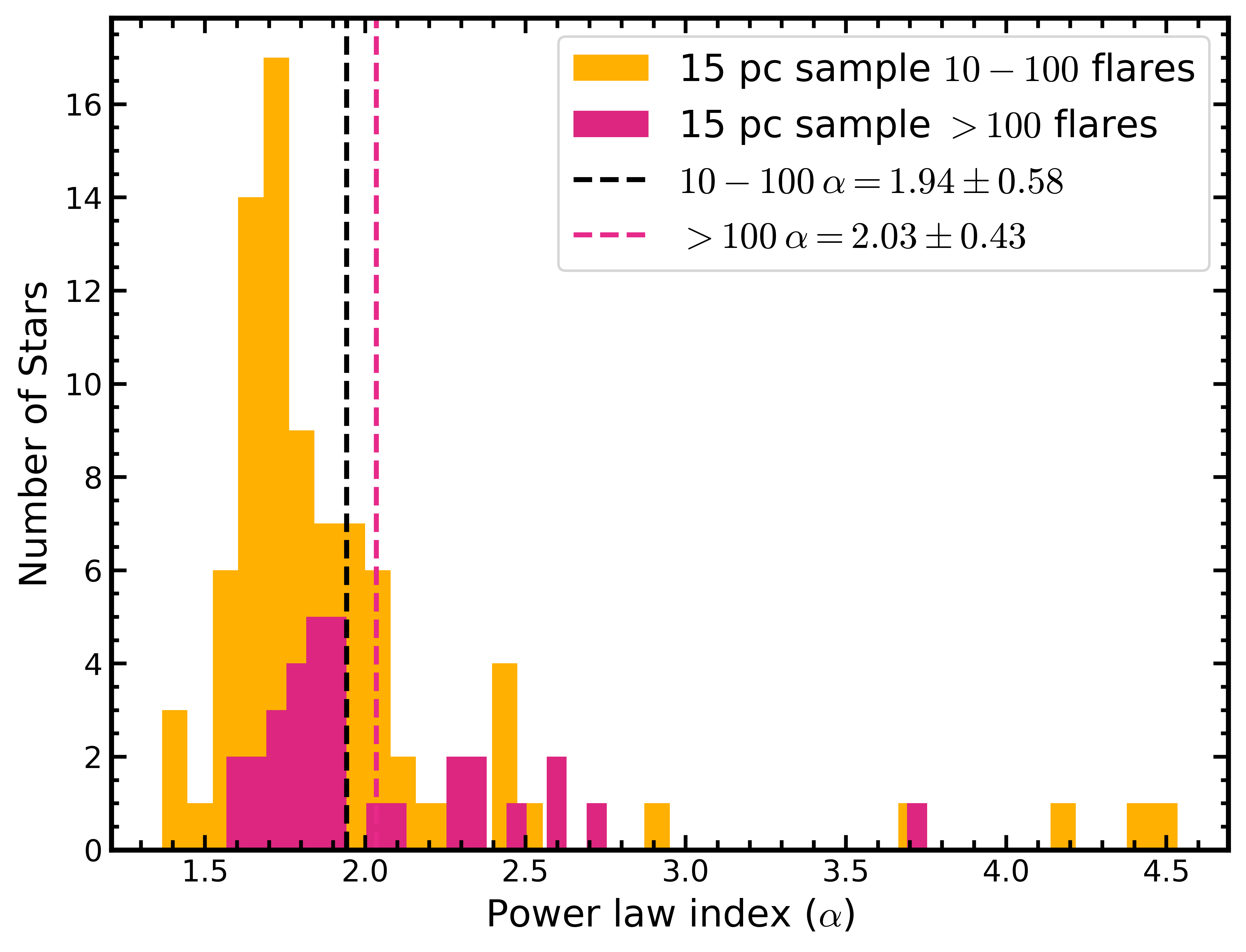}
    \caption{Histograms of the power law exponents ($\alpha$ values) fit to the flaring M dwarfs from the Gaia DR3, volume-limited 15 pc sample. We detected flares in the \tess{} observations for 276 of 538 targets. As with Figure \ref{fig:slopes}, we only include the $\alpha$ values for stars with $>10$ flares (117 stars), \dif{but due to the smaller sample size, we show separate distributions for} those with 10--100 flares (84 stars) in gold, and those with $>100$ flares (32 stars) in pink, rather than splitting by relative activity. The vertical dotted lines mark the average $\alpha$ values of $1.94\pm0.58$ and $2.03 \pm 0.43$ respectively. While similar to the total sample that includes the \cite{Gunther_2019} M dwarfs, we find larger averages and larger standard deviations for these exponents. We observe a smaller spread of $\alpha$ values for FFDs containing more flares.}
    \label{fig:15pc}
\end{figure}

In Figures \ref{fig:slopes} and \ref{fig:15pc} we show histograms of the best-fit $\alpha$ parameters from our MCMC power law fits on stars with more than 10 detected flares within highly active sample. We also investigate whether this average $\alpha$ changes for highly active stars with a larger number of detected flares ($10-100$ flares compared to $>100$ flares) because studies have reported improvements on $\alpha$ uncertainties and power law fits for stars with larger number of flares \citep{AltaiPony,Aschwanden_2021}. Figure \ref{fig:slopes} represents the highly active stars from the combined \cite{Gunther_2020} and volume-limited 15 pc sample, and Figure \ref{fig:15pc} shows the highly active stars from only the volume-limited 15 pc sample.
In Figure \ref{fig:slopes} we show the $\alpha$ values for two bins of the highly active category: FFDs fit with 10--100 detected flares as the gold distribution (\dif{278} stars), and those fit with $>100$ flares in pink (86 stars). For the distributions of the 15 pc sample $\alpha$ values (Figure \ref{fig:15pc}), we found 84 M dwarfs with 10--100 flares (gold), and 36 stars with $>100$ flares in their FFDs. The vertical blue line in Figure \ref{fig:slopes} indicates the best-fit $\alpha$ parameter for the combined FFD of the lower activity stars computed in Figure \ref{fig:mcmc}. The uncertainty for the lower activity exponent in Figure \ref{fig:mcmc} is based on the standard deviation of the MCMC posteriors, and is thus different from the uncertainties we report for the distribution highly active $\alpha$ values, which are instead the standard deviation of $\alpha$ values computed across FFDs of multiple stars. \cite{Aschwanden_2021} suggests binned power law fitting methods allow for the uncertainty on the best-fit power law exponent can be estimated by 
\begin{equation}
\sigma_\alpha =\frac{\alpha}{\sqrt{n_{flares}}}
\label{eqn:uncert}
\end{equation} with $n_{flares}$ representing the number of flaring events making up the fitted range of a flaring distribution. For the lower activity sample, this gives a slightly larger uncertainty of \dif{0.068} for the lower activity power law, which consists of \dif{858} flaring events. We include this uncertainty estimate in Figure \ref{fig:slopes} to better compare with the average and standard deviations of the highly active $\alpha$ distributions.

\section{Results}
\label{sec:results}
In the JWST Cycles 1--3 sample of 41 M dwarfs, we detect flares in the \tess{} light curves of 14 targets, finding a total of 185 flares. In the total sample containing all M dwarfs within 15 pc from Gaia, and the additional flaring M dwarfs from \cite{Gunther_2020}, we detect a total of 50,730 flares.

Figure \ref{fig:FFD} shows the FFDs for the individual JWST targets (differentiated by color and shape) in comparison to the FFDs of the M dwarfs cataloged by \cite{Gunther_2020}, and the FFDs of our volume-limited 15 pc sample from Gaia DR3 (blue and gold contours). The cumulative flare frequency and ED measurements we obtain for the JWST Cycles 1 and 2 targets (left) largely fall in the lower activity flaring region, with the exception of the highly energetic flares on TOI-2445, as well as popular JWST targets TRAPPIST-1 and LTT 1445. The FFDs \dif{we compute} for the JWST Cycle 3 targets (right) are within the highly active flaring category, except for GJ 1151 which has been reported to have a generally lower magnetic activity level, long rotation period and infrequent flaring \citep{Pope_2021,GJ1151_Blanco-Pozo_2023}. We expect targets that fall in the highly active region will have increased levels of stellar activity contamination to contend with when analyzing their planets' transmission spectra. 

Figure \ref{fig:slopes} shows the distribution of power law exponents we calculate for the highly active FFDs across both samples. We show a distribution of values for stars with \dif{more than 10 detected flares (gold), which has an average power law exponent $\alpha = 1.81$ and a standard deviation of $0.32$, indicated by the pink dashed line}. Due to the asymmetric tail of \dif{this distribution}, and confirmed via Anderson-Darling tests, we reject the null hypothesis that these $\alpha$ parameters follow a normal distribution \citep{Anderson-Darling}. The exponent of our combined lower activity FFD ($\alpha =$\lowactivity) is near the average 10--100 flare $\alpha$ value, as indicated in this Figure by the vertical blue line, with the distribution and best fit power law shown in Figure \ref{fig:mcmc}. The combined FFD of the lower activity stars consists of \dif{858} flares from \dif{170} different stars. The \dif{blue shaded region and} uncertainty for $\alpha$ (0.078) for the lower activity FFD in Figure \ref{fig:slopes} is the estimated error based on Equation \ref{eqn:uncert}, described in \cite{Aschwanden_2021}. This provides a better error estimate from the fitting routine, by incorporating the total number of flaring events within the fitted range.

In Figure \ref{fig:15pc}, we show the $\alpha$ values fit to the FFDs of stars from the volume-limited 15 pc sample alone. These histograms represent exponents from 117 M dwarfs with $>10$ flares, from the 276/538 stars that we detect flares within the \tess{} observations of the entire sample. We split this sample \dif{to show }$\alpha$ values for stars with 10--100 flares in gold, and those with $>100$ stars in pink. We find and mark average values of $1.94\pm0.58$ and $2.03\pm0.43$ with dashed vertical lines for these sub-samples respectively, where uncertainties on the averages are the standard deviations of each distributions. We observe the same extended tails as in the combined samples, with larger standard deviations. Both of these average values are higher than those of the combined sample, but are consistent within uncertainties and with the lower activity power law exponent as well. We note that we did not cut by relative activity level as we did for the total sample, due to the smaller sample size. 

We find that the distributions for stars with 10--100 flares and those with $>100$ flares appear largely the same. There is some deviation in the tails that extend to higher $\alpha$ values, though we caution interpreting these as different populations due to the low number of stars in each distribution. We don't see large difference of $\alpha$ fit to the FFD of the combined lower activity stars, compared with the average values of the highly active samples. 
\dif{The uncertainties we report in both figures and these results are either an estimate based on the size of the distribution in the case of the lower activity sample (Eqn. \ref{eqn:uncert}), or are the standard deviations of values within each highly active star sample.} These uncertainties therefore may not be representative of errors introduced by assuming the power law function we fit is correct, or variation of the parameters based on the data we selected to fit for the highly active stars. This is problematic for stars such as AD Leo, which is better fit by either a broken power law or two different power laws, with one half showing very shallow decline, followed by a steep distribution for its higher energy flares. Though less drastic, AU Mic is another target that shows broken power law behavior in its FFD, shown by the purple diamonds in the right panel of Figure \ref{fig:FFD}. The flares with ED $>10^2$ have a higher cumulative frequency than is indicated by the power law trend of its lower energy flares.

\section{Discussion}
\label{sec:summary}
\subsection{Literature comparisons}
\label{sec:litcomp}
\subsubsection{Flare catalog completeness}
\label{sec:completeness}
\dif{Our flare detections for the volume-limited sample of M dwarfs and those we detected on the M dwarfs targets provided by \cite{Gunther_2020}, can be contextualized within the literature on flaring statistics derived from optical photometric observations. Long- and short-cadence observations by both the Kepler and \tess{} missions have provided numerous detections and characterizations of flaring on stars spanning spectral types A--M. While the results of these works have evolved slightly based on methodology developments, there is agreement between studies on key findings and limitations that we have seen in this work as well.}

\dif{Limitations of data quality and time-sampling appear to be the driving factor in discrepancy between studies. Kepler offered both long-cadence (LC; 30-minutes) and short-cadence (SC; 1-minute) observations, where \tess{} has offered 2-minute and 20-second cadences. Findings from early studies using Kepler \citep[e.g,][]{Davenport_2014,Balona_2015, Davenport_2016, VanDoorsselaere_2017,Yang_2017} largely used the long-cadence observations for flare detections and statistics.} 
\dif{After providing a direct comparison of flares in Kepler LC and SC data, \cite{Yang_2018} suggested properties like flare energy and amplitudes are typically underestimated from LC detections, while flare durations are overestimated when compared to detections in SC data. They also conclude that approximately $\sim 60\%$ of SC flare detections are lost by LC Kepler data and emphasize caution when interpreting previous studies results on flaring star fractions and flaring frequency. }

\dif{We investigate the overlap of our flare detections with other literature studies that have made their flare catalogs publicly available, by comparing each studies reported flare start and stop time, or their flare peak time if the full durations are unavailable. The studies with publicly available flare catalogs from \tess{} data include: \cite{Gunther_2020,Pietras_2022,Howard_2022,Yang_2023,Feinstein_2024,Lin_2024,Zhang_2024,Seli_2025}. After determining the stars our catalogs have in common by TIC ID, we compare the flares we detect for each star. For the catalogs with reported flare start and stop times, we identify whether this flare duration overlaps with our flare durations for the same star. For those only listing peak times, we check if their peak times fall within one cadence of our flare duration (start and stop times). The tolerance of one cadence is included to account for the rounding of peak times by the available catalogs. }

\dif{\cite{Gunther_2020} offered one of the first flare studies using \tess{} 2-minute cadence data from sectors 1 and 2. Using \aP{}, we detect flares in all of the 685 M dwarfs they included in their catalog. With their publicly available flare catalog, we calculate the overlapping detections in our samples to be 82.03\%. This comparison is based on their reported flare peak times for the stars we had in common, limited to the \tess{} sectors they investigated after cutting for binary star systems.}

\dif{We have 570 stars in common with the catalog of \cite{Pietras_2022}, and measure a 69.45\% overlap with their flare catalog based on their reported peak times. Similarly, we find overlap of 48.06\% for the flares detected by \cite{Feinstein_2024}, based on their peak times, though this was limited to only 18 stars we had in common with their sample. We find agreement of 83.44\% of the flares detected for 481 stars we have in common with the catalog presented by \cite{Seli_2025}. We find the largest agreement with \cite{Yang_2023}, recovering 88.86\% of the flares they detected on 558 stars we had in common. We used their provided flare start and stop times to find matches based on the overlap between our flare candidates. Using the same criteria for matching, we find overlapping flares for 631 stars in common with \cite{Lin_2024}, where we recovered 71.62\% of the same flares for these stars.}

\dif{In general, we find less consistent flare matching with studies using 20-second cadence data rather than 2-minute cadence data \citep{Howard_2022,Zhang_2024}. This could be due to selection effects based on detection methods, which limit 2-minute cadence surveys to flare durations $>$ 6 minutes, based on the three consecutive outlier detection criteria. Additionally, as mentioned by \cite{Zhang_2024}, the higher photometric noise of 20-second cadence data means some lower amplitude detections were missed in their comparison with \cite{Yang_2023}. }

\begin{figure}
    \centering
    \includegraphics[width=0.98\linewidth]{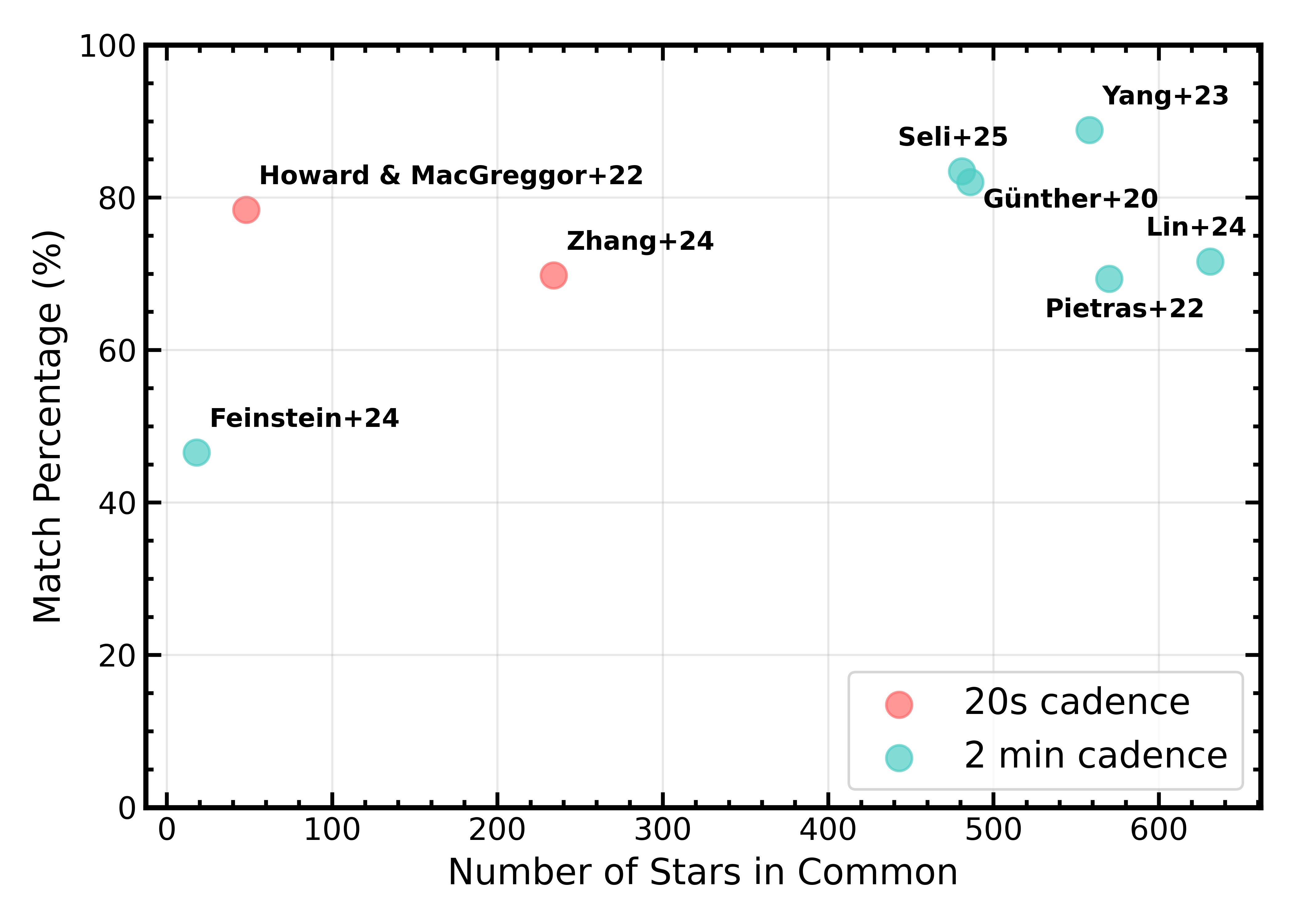}
    \caption{\dif{The percentage of flares matched plotted against the number of stars in common with our catalog for the studies with publicly available catalogs. These studies include: \cite{Gunther_2020,Pietras_2022,Howard_2022,Yang_2023,Feinstein_2024,Lin_2024,Zhang_2024,Seli_2025}. The points are colored based on the cadence of the observations, with red representing 20-second cadence studies, and teal representing 2-minute cadence studies. Each point is annotated with the abbreviated reference for the corresponding study.}}
    \label{fig:catoverlap}
\end{figure}

\dif{Figure \ref{fig:catoverlap} shows the results of these comparisons, plotting the flare detection match percentage against the number of stars we have in common with each of the publicly available \tess{} flare catalogs. Color indicates the cadence of the \tess{} data used in each study, with red representing 20-second data and teal representing 2-minute cadence. We see that typically our methods find more flaring stars in common with the 2-minute cadence surveys than the studies using 20-second data, though flare match percentage is fairly consistent on average.} 

\dif{We find that a large portion of the differences in flare detection between our study and the other studies presented here are due to the observational cadence of our selected data as well as the S/N of the ``missed" flares, with undetected flares preferentially occupying areas of $<$50\% recovery in our injection recovery tests. We also note that another contribution to the mismatch in catalogs is due to potential confusion in single-counting multiple flares or double counting single flares with complex shapes \citep{Davenport_2014}.}

\subsubsection{Flare frequency distributions and power law exponents}
\label{sec:ffdPLs}
An explanation proposed for the power law relation between stellar flare frequency and flare energy is called self-organized criticality \citep[SOC; ][]{Bak_1987,Bak_1988,Aschwanden_2021,Feinstein_2022}. This theory says that when the magnetic field of the star reaches a critical instability, local perturbations can trigger flare energy release via magnetic reconnections, which subsequently trigger smaller perturbations and flares that will follow the energy power law distribution. For FFD power laws, $\alpha = 2$ is suggested to be the critical value that determines the total energy of the distribution. This theory has yet to be proven however, and there are other mechanisms in which power laws can originate for cumulative distributions \citep{Newman_2005}.

\dif{Other studies \citep[e.g.,][]{Mullan_2018,Gao_2022} investigate instances of the observed broken or two-component power law shape, like those we see most clearly for AD Leo standing out above the highly active FFDs in the right panel of Figure \ref{fig:PLsplit}, but in many other stars as well (see Figure 3 of \citealt{Vida_2024} for another example of a two-component flare frequency distribution). Some state that this behavior is indicative of a physical process \citep{Mullan_2018} or an inherent limit on the flare energy that a certain star can produce. The other explanation is that this occurs due to an observational biases related to incompleteness of low energy flares at the left of the distribution due to a lack of photometric precision \cite{Gao_2022}, or potentially the rarity of high energy events towards the right of the distribution, and the scarcity of data available to characterize them. Some studies have suggested that both may contribute to deviations from the ideal power law and that this does not contradict with the SOC power law relationship \citep{Aschwanden_2015}. Whether this is a result of observational biases, physical mechanisms or both is still  unconfirmed, but it does appear for some of the stars in our sample, and therefore introduces some uncertainty and outliers in our power law exponent comparisons. While we try to roughly select the middle portion of each FFD by excluding the lower 20\% of energies, this cannot fully account for the most drastically bimodal FFDs, or those with heavily flattened low flare energy distributions that extend beyond this cut. }

\begin{figure*}
    \centering
    \includegraphics[width=0.75\linewidth]{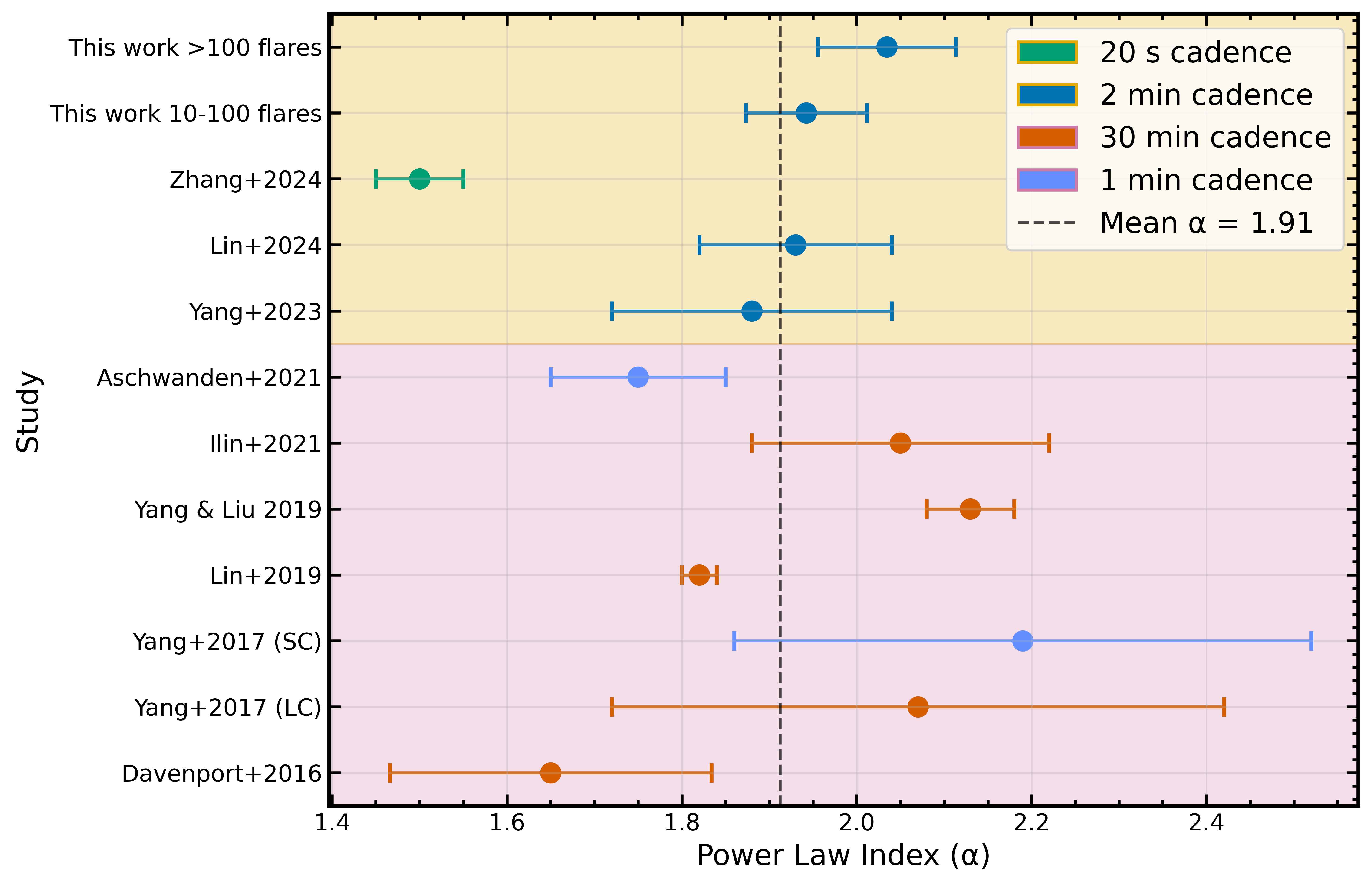}
    \caption{\dif{A comparison of power law exponents ($\alpha$) fit to M-type star flare frequency distributions, from various studies using Kepler and TESS data. The points represent the best-fit $\alpha$ values with error bars indicating uncertainties. The dashed line marks the mean $\alpha$ across all studies. Background shading differentiates between Kepler (magenta) and TESS (yellow) studies, while point colors indicate observation cadence: 20 seconds (green), 2 minutes (blue), 30 minutes (orange), 1 minute (purple). We find that our work is consistent with the average power law index of $\alpha$ = 1.91 for both highly active stars with more than 100 detected flares as well as active stars with between 10 and 100 flares. Our low activity stars have an active power law slope of 1.99 $\pm$ 0.07, broadly consistent with the range of measurements in other flare studies.}}
    \label{fig:alphas}  
\end{figure*}
Our computed power law exponents fall within the expected range of those found by other studies. \dif{\cite{Davenport_2016} created a catalog of flaring stars from Kepler both long- and short-cadence data, producing FFDs for individual stars. Transforming the results of their power law fits to the form of Equation \ref{eqn:pl}, they report an average $\alpha$ value of 1.64 with a standard deviation of 0.184 for M-type stars (M$<$ M$_\odot$). Due to their flare threshold requirements before and consistency of FFD computation between cadences, these exponents should be accurate despite potential inaccuracies in the actual flare properties due to long cadence detections described by \citep{Yang_2018}.} 

\dif{\cite{Yang_2019} published a LC flare catalog from all of Kepler after improving detection and validation methods to address some of the challenges described in \citep{Yang_2018}. They included FFDs and power law fits by spectral type finding $\alpha \sim 2.13 \pm 0.05$ for M-type stars and $\alpha \sim 2.09 \pm0.10$ for fully convective M-types ($T_{\mathrm{eff}}<3200$ K).} \cite{Aschwanden_2021} \dif{compiled all $\alpha$ values from pre-Kepler through Kepler flare studies predating it (See their Tables 1 and 2 of the values and references), for which they found an average of $\alpha = 1.99 \pm 0.35$ for M-type stars from the reported values of 19 different Kepler flare datasets. }


They also show how the range of $\alpha$ values were also tighter constrained with a smaller standard deviation for larger sample of flares: $1.5 \lesssim \alpha \lesssim 3$ for the full sample vs $1.9 \lesssim \alpha \lesssim 2.3$ for distributions with $>1000$ flares. We see this in our results as well (Fig. \ref{fig:15pc}), with the standard deviations on our $\alpha$ calculations decreasing for stars with greater number of flares. \dif{For optical wavelengths, when including only samples with a small number of flares, \cite{Aschwanden_2021} found  $\alpha_{OPT}=1.93\pm0.40$, compared to $\alpha_{OPT}(n_{flares} >1000)=2.09\pm0.24$ which they report for exponents fit to larger samples.} \dif{The FFDs we computed for the 15 pc sample alone (Fig. \ref{fig:15pc}), follow these findings closely, with our results for stars with a smaller number of flares (10-100) and those with $>100$ flares, having average values of 1.94 and 2.03 respectively.}

\dif{In addition to our agreement with these compiled values from Kepler flare datasets, within uncertainties, all of our average $\alpha$ values are in agreement with the literature review values shown in \cite{AltaiPony}. Their review includes some of the same Kepler studies analyzed by \cite{Aschwanden_2021}, but they also incorporated $\alpha$ values fit to FFDs from other optical flare surveys, both ground-based \citep[e.g,][]{Hilton_2011,Howard_2019}, as well as early \tess{} flare datasets \cite{Gunther_2020,Feinstein_2020}. Figure 13 of \cite{AltaiPony} presents the distribution of these published $\alpha$ values.} The majority of studies on flare statistics they included also seem to converge around $\alpha \sim 2$. The $\alpha$ values calculated in their study ranged from 1.84 to 2.39, but were better constrained to $\alpha \sim 1.89-2.11$ when accounting FFDs consisting of a larger number of flares ($n_{\mathrm {flares}}>50$). 

\dif{Our average $\alpha$ values also align} well with some studies focused on FFD correlation with effective temperatures and spectral types. \dif{Early large survey (Kepler) flare studies investigating power law exponents by spectral type found consistent values for F--K stars, with outliers for A-type stars \citep[e.g,][]{Balona_2015,Aschwanden_2015,Davenport_2016,Pedersen_2017,Yang_2017,Yang_2019}.} \cite{Lin_2019} found $\alpha$ values of 1.82 and 1.86 for M and K dwarfs in K2 observations. In more recent work, \cite{Yang_2023} search the first 30 TESS sectors for flares, finding M- and K-type $\alpha$ values to be 1.85 and 1.81 when excluding binaries in close agreement with our findings.

In a much larger survey of all \tess{} light curves using multiple machine learning models, \dif{\cite{Lin_2024}} computes an average exponent $\alpha= 1.93 \pm 0.11$ for the combined FFD of all M dwarfs in \tess{}, which is in strong agreement with our averages (Figs. \ref{fig:slopes} \& \ref{fig:15pc}). However, their samples only account for literature confirmed eclipsing binaries (EBs) rather than performing any cuts with the Gaia information as we did. Their $\alpha$ values for the A-K spectral types are smaller than M-types, contrary to their previous findings \citep{Lin_2019}, and some of these are less consistent with prior studies. They do perform a more rigorous review of the sensitivity for the lower energy detections than we can perform \aP, and credit discrepancies with previous studies being due to the different energy ranges being chosen to fit with the power law. 

\dif{Figure \ref{fig:alphas} shows the $\alpha$ values fit to only M-type stars from literature studies that analyzed either Kepler or TESS data. Uncertainties are their reported values, which do vary slightly in computation method. We use bootstrapping to find the mean values of our 15 pc sample, using the individual fit $\alpha$ values and their uncertainties, to obtain a refined uncertainty of the peak value, which are 0.069 and 0.079, for the stars with 10--100 flares and those with $>100$ flares, respectively. Rather than using our distribution standard deviations reported in Figure \ref{fig:slopes} as error bars, we find these errors on our central values are more comparable to the reported uncertainties of these studies, which largely fit multi-star FFDs instead of individual FFDs as we did in this work. Sorted by publication date, we shade the region behind the scatter points as magenta for the studies using Kepler data, and yellow for the studies using \tess{} data. We indicate the cadence of the data used for each study by the color of each data point and error bar, including both long (LC; 30-minute) and short cadences (SC; 1-minute) for Kepler, and both 2-minute and 20-second \tess{} cadences.}

\dif{We find our $\alpha$ values to be mainly consistent with literature findings for M spectral types, especially between studies using TESS data. \cite{Zhang_2024} does describe some selection effects, which we discussed in Section \ref{sec:completeness}, which may contribute to their reported $\alpha$ falling further outside the typical values of \tess{} studies.}


FFD power law behavior and correlations with spectral type, rotation periods, stellar activity indicators, and other stellar evolutionary traits is still an area of investigation. As detection and computation methods continue to improve, we may be able to determine these, though all results appear to be heavily dependent on flare detection completeness at the lower energy end, which no current photometric instrument or detection methodology can adequately address. \dif{Biases in detection methods and data cadence have been observed to affect flare properties and statistics, so these factors need to be addressed to clarify the results and determine the origins of the observed break in the power laws of flaring frequency distributions. Future missions such as PLATO \citep{Rauer_2025}, with improved photometric precision compared to \tess{} can alleviate observational biases related to detecting low amplitude flares. Continuous and multi-year coverage similar to Kepler will allow for more detections of larger energy and rarer events than \tess{} can with its 27 day sectors, which can reduce the truncation at the higher energy end of flare frequency distribution power laws. The shorter observation cadence modes of PLATO will offer advantages over both \tess{} and Kepler in detecting lower energy flares of shorter durations, and to constrain flare properties that can be lost in longer cadence data \citep{Howard_2022,Zhang_2024}.}

\subsection{Relevance to JWST Planetary Atmosphere Studies}
\label{sec:JWST}
\cite{Rackham_2018} defines the transit light source effect, which describes how observed transmission spectra of planets will be imprinted with any spectral difference between the illuminating light source and the disk-integrated stellar spectrum, specifically the differences due to starspots and faculae. 

Studies citing contamination due to the transit light source effect are generally more focused on slowly varying noise sources such as spots. To understand the larger picture of the various stellar phenomena contaminating transmission spectroscopy results, however, \cite{Howard_2023} investigated near-infrared spectra of flares on TRAPPIST-1 during JWST transit spectroscopy. They state that flares must be considered a source of unpredictable contamination, and found flares large enough to impact transit spectra occur at a rate of  $3.6^{+2.1}_{-1.3}$  flares day$^{-1}$ for TRAPPIST-1. 

We find a lower frequency for TRAPPIST-1, though only detected one high energy flare in its only available \tess{} light curve where we are unable to detect the more frequent, lower amplitude flares. At a \tess{} magnitude of 13.85, the noise of this target is very high, as discussed in Section \ref{sec:I-R}, meaning \aP{} can only detect high energy flares. \cite{Howard_2023} do say that up to 80\% of flare contamination can be removed from transmission spectra via non-local thermodynamic equilibrium (non-LTE) radiative hydrodynamic modeling and subtraction, with mitigation most effective from 1.0-2.4 $\mu\mathrm{m}$. Due to their successful removal from the TRAPPIST-1f and b transmission, they suggest transits affected by flares may still be useful for atmospheric characterization efforts. One Cycle 4 program (JWST-GO-7068) will examine these prospects further by targeting 5 active M dwarfs (TIC 261560580, TIC 231020638, TIC 80427281, TIC 89502706, and TIC 207140650) to measure flaring in the near infrared spectra. Stellar flares are linked to the total magnetic activity of a star, however, and can indicate the presence of other stellar heterogeneities, including those which are more difficult to mitigate \citep{Notsu_2013, Doyle_2020,Feinstein_2020}. 

These more difficult features did heavily influence the retrieval of the transmission spectra of planet TRAPPIST-1 b presented by \cite{Lim_2023}. They deemed their results inconclusive due to the stellar contamination, and suggest that work to characterize and disentangle stellar activity from JWST transmission spectra must be done to overcome these challenges for future explorations. Looking at the next planet in the system, TRAPPIST-1 c, \cite{Radica_2024} also found that their retrievals could be entirely explained by stellar contamination for both visits taken 367 days apart. \cite{Moran_2023} encountered similarly inconclusive results in their JWST atmospheric retrieval of GJ 486 b, in which they found nearly equal agreement between a water-rich atmosphere scenario and a transmission spectrum contaminated by cool unocculted star spots. \cite{May_2023} presented the results of the JWST observations of GJ 1132 b, for which we found two previously unreported flares in its \tess{} data. In their work, they state that data that are inconsistent from epoch to epoch, suggesting that variations in stellar activity may be confounding their ability to probe the atmosphere of GJ 1132 b. 


Two NIRSpec transits of LHS 1140 b have been observed
and were published by \cite{Damiano_2024}, in which they suggest a scenario of an N$_2$- or CO$_2$-dominated atmosphere is favored over the predicted scenario of a H$_2$ rich atmosphere due to a lack of observed CH$_4$ or CO$_2$ transmission spectral features. Their study was also impacted by stellar activity, but they say the model with stellar heterogeneity included does not significantly influence their result. They recommended a future campaign to observe 9 additional transits (over multiple years due to the 25 day orbital period of the temperate planet) to detect a CO$_2$ absorption feature to disentangle their possible atmospheric scenarios, but warn that such a program could be hindered by stellar activity that can vary from visit to visit. This activity is consistent with previous attempts to observe the planet's atmosphere with the HST WFC3 observations from \citep{Edwards_2021}, where transit depth changes in the 1.1–1.7 $\mu$m wavelength range could be reasonably fit by both atmospheric models and stellar spot contamination models. More recently, \cite{Cadieux_2024} published the NIRISS transmission spectrum of LHS 1140 b, showing strong evidence for unocculted faculae covering $\sim20\%$ of the star. They rule out the same scenario as the NIRSpec authors, and recommend several years worth of consecutive NIRISS/NIRSpec observations to properly characterize stellar contamination and detect any potential secondary atmosphere. While we detected no flares in the 2 extant \tess{} Sectors of LHS 1140, this M dwarf still exhibits activity that complicates transmission spectroscopy observations at some level. 

We did not detect any flares in the light curves of TOI-270, whose JWST observations of its planets b and d
were recently published by \cite{Holmberg_2024}. They find evidence for signatures of CH$_4$ and CO$_2$, with little evidence of stellar heterogeneities in either planet's spectra, which supports the conclusion of the detection paper by \cite{Gunther_2019} that this is a quiet M dwarf. \cite{Gunther_2019} found no evidence of stellar flares or activity in the first three \tess{} Sectors it was observed in (3,4,5), which is confirmed in this study, and our non-detections within the more recent Sectors, 30 and 32.

Not only complicating JWST observations, \cite{Barclay_2025} has submitted their HST (Wide Field Camera 3) observations for the L 98-59 c, for which we detected 5 flares in the star's \tess{} observations. Their results are inconclusive as they cannot rule out a scenario where a signal they observe originates from stellar heterogeneities. They are hopeful that HST and JWST observations in the near future can provide a more conclusive scenario. 

Emission spectroscopy, and more specifically secondary eclipse spectroscopy, offers an alternative and promising approach for investigating M dwarf planet atmospheres \citep{Redfield_2024}. Tidally locked terrestrial planets, such as habitable zone M dwarf planets, have a dayside temperature that is regulated by the planet's atmosphere \citep{Joshi_1997}. Although the observational S/N is too low to identify individual features for smaller and cooler planets compared to emission spectra of gas giants, the presence of such an atmosphere can be indirectly inferred by observing the thermal emission during the planet's secondary eclipse \citep{Belu_2011,Morley_2017,Koll_2019,Kreidberg_2019,Crossfield_2022,Zieba_2023}. 

The recently published near- to mid-infrared JWST transmission spectrum of L 168-9 b \citep{Alam_2024} shows a lack of atmospheric features. They suggest that 15 $\mu$m MIRI eclipse observations could discern between a high mean molecular weight and no atmosphere scenarios for the planet. A similar recommendation to observe eclipses with MIRI was given by \cite{Lustig-Yaeger_2023}, who presented the NIRSpec observations of two transits of LHS 475 b, a tidally locked rocky planet which had another featureless transmission spectrum. These authors found no evidence for stellar contamination by spots or faculae in their observations, and we found one flare in the \tess{} data. 

Although the transmission spectroscopy observations of LTT 1445A b (JWST-GO-2512) have not yet been published at the time of this work, \cite{Wachiraphan_2024} has published the results of their JWST MIRI/LRS thermal emission analysis on three observed secondary eclipses of the planet. Their results disfavor the presence of a thick atmosphere, instead favoring a model of instant reradiation from a rocky surface. They suggest this could be due to atmosphere erosion or escape, which is well supported by the recent flaring and activity study of this star (as well as Cycle 1 target GJ 486) by \cite{Diamond-Lowe_2024}. We detected 15 flares in the \tess{} data for LTT 1445, shown in the left panel of Figure \ref{fig:FFD}, on the edge of the highly active category. Though it was pointed out by \cite{Diamond-Lowe_2024}, that the \tess{} light curves are contaminated by an active binary companion 7" away that could also be the origin of these flares. 

However, their analysis revealed higher flaring in UV and X-ray wavelengths using HST and \textit{Chandra} observations, despite long rotation periods and indicators of low activity in the optical. The conclusions of their study, as well as previous works \citep[e.g.,][]{Loyd_2018, Jackman_2024} suggest that optical activity indicators cannot predict the high energy activity that powers processes such as thermal atmospheric escape \citep{Lammer2003} and perturbation of thermochemical equilibrium \citep{Hu_2012}. The implications of the total activity of a star not mirroring the optical activity is supported by some of our results as well. Where we see no flares in the optical \tess{} observations of some targets that have possible contamination in their planets' IR transmission spectra, such as L 168-9, LHS 1140, GJ 486.

Flares also release large amounts of energy in the higher-energy regimes. While these flares may not show up in the JWST transmission spectra or directly contaminate the IR observations, their presence, along with persistent XUV stellar flux, could impact the atmospheres and habitability of orbiting exoplanets over an M dwarf's long lifespan, via photoevaporation, erosion or ozone depletion \citep{owenphotoevap,Segura_2010,Tilley_2019,Santos_2023}. The inverse is also true, in that low flare frequencies may result in UV radiation that is insufficient to power disequilibrium chemistry processes \citep{Ranjan_2017,Rimmer_2018}. Therefore, a star's flare frequency in various wavelengths can inform studies on the presence or retention of its planet's primordial \citep{Kasting_1983,Owen_2012,Airapetian_2017,Rogers_2021}, or secondary atmosphere \citep{Dong_2018,Kite_2020}.

Optical activity indicators however, can inform studies of host-star spot coverage and spot crossing at time of transit, so performing simultaneous or near-simultaneous optical observations with the IR transmission spectroscopy could be the best way to inform these studies of the origins of any observed spectral features.

\subsection{The flare-active star AU Mic}
\cite{Gilbert_2022} performed a thorough search of the \tess{} data of AU Mic, a planet-hosting pre-main-sequence M dwarf that shows large photometric variability due to the high amount of stellar activity. In their study they analyze the flaring activity and spot modulation of the star in the 2 minute cadence data for sectors 1 and 27, as well as the 20 second cadence data in sector 27. While we detect more flares in the 2 minute cadence light curves alone (149 compared to their 98), our flare detection is largely in agreement with their results, as the total number of flares between the two sectors increases to 173 when they used the 20 s data for Sector 27. They report more detections of lower amplitude flares in 20 s data than was possible in the 2 minute cadence data, however the 2 minute cadence SPOC light curves provide us less scatter, which is more suitable for the \aP{} detection criteria (as shown in Section \ref{sec:I-R}).

Approved JWST Cycle 3 GO program 5311 will retrieve the NIR transmission spectra of AU Mic b, with simultaneous FUV observations by HST. With these simultaneous observations, the investigators of this program hope to develop tools to mitigate flares from NIR transmission spectra, as well as provide a unique opportunity for multi-wavelength characterization of stellar contamination.

\section{Conclusions}

In this work, we performed a large-scale search for flares in the \tess{} light curves of M dwarfs, focusing on both a volume-limited 15 pc sample and planet-hosting targets scheduled for JWST transmission spectroscopy. By detecting and characterizing over 50,000 flares across nearly 500 stars, we computed flare frequency distributions (FFDs) and fit power laws to analyze trends in stellar activity.

We found that the power law exponent $\alpha$ remains consistent across different activity levels that we defined as relatively lower and higher activity M dwarfs. The combined FFD of the lower activity stars yielded a best-fit $\alpha$ of \lowactivity, while the higher activity stars were best constrained to values between $1.6<\alpha< 2.46$ for both samples of known flaring M dwarfs, and all M dwarfs within 15 pc. Even among the more active stars, the $\alpha$ distributions showed minimal deviation, which could suggest that the underlying physics driving flare production may be largely uniform across the M dwarf population. 

We showed that many JWST M dwarf targets fall within the same activity regimes as the broader 15 pc sample, offering critical context for interpreting transmission spectra. While some targets, like TRAPPIST-1 and TOI-2445, exhibit high energy flaring, the majority display more modest activity, suggesting that stellar contamination in upcoming JWST spectra may vary significantly from target to target. Some targets that did not show flaring in their \tess{} observations, or that were undetectable with our methods, have still had contaminated follow-up observations, emphasizing the importance of characterizing the various forms of stellar activity for these studies. 
\begin{acknowledgements}
We thank the anonymous referee for their comments and suggestions for this manuscript. Their input significantly improved this manuscript. 

Some/all of the data presented in this paper were obtained from the Mikulski Archive for Space Telescopes (MAST) at the Space Telescope Science Institute. The specific observations analyzed can be accessed via \citep{MAST_TESS}. STScI is operated by the Association of Universities for Research in Astronomy, Inc., under NASA contract NAS5–26555. Support to MAST for these data is provided by the NASA Office of Space Science via grant NAG5–7584 and by other grants and contracts.
This research has made use of the SIMBAD database, operated at CDS, Strasbourg, France.
The authors acknowledge UFIT Research Computing for providing computational resources and support that have contributed to the research results reported in this publication. URL: \url{http://www.rc.ufl.edu}
\end{acknowledgements}

\facilities{\tess, \Gaia{} \citep{GaiaDR3}, Mikulski Archive for Space Telescopes \citep{MAST},  } 

\software{\texttt{AltaiPony} \citep{AltaiPony,Davenport_2016}, \texttt{astroquery} \citep{astroquery}, \texttt{corner.py} \citep{corner}, \texttt{edmcmc} \citep{vanderburgedmcmc}, \texttt{Lightkurve} \citep{lightkurve}, matplotlib \citep{matplotlib}, \texttt{Scipy} \citep{Scipy}} 
\vspace{5mm}
\clearpage

\bibliography{bibli.bib}

\end{document}